\documentclass{jfm}
\usepackage{epsfig,graphics,graphicx,amssymb,amsmath, color,natbib,upmath}

\usepackage[sfdefault=cmbr]{isomath}

\def\e{\epsilon}
\def\v{\vectorsym}
\def\t{\tensorsym}
\def\bcdot{\boldsymbol{\cdot}}

\include{help}
\title[The sedimentation of flexible filaments]{The sedimentation of flexible filaments}

\author[L.~Li, H.~Manikantan, D.~Saintillan and S.~E.~Spagnolie]%
{Lei Li$^{1}$, Harishankar Manikantan$^2$, David Saintillan$^2$ \\ and Saverio~E.~Spagnolie$^1$\thanks{Email address for correspondence: spagnolie@math.wisc.edu}}

\affiliation{$^1$Department of Mathematics, University of Wisconsin-Madison, 480 Lincoln Drive,  \\
Madison, WI 53706, USA\\[\affilskip]
$^2$Department of Mechanical Science and Engineering, University of Illinois at Urbana-Champaign, 1206 West Green Street, Urbana, IL 61801, USA}

\date{\today}

\pubyear{2013}
\volume{??}
\pagerange{??--??}
\date{\today}
\usepackage{graphicx,color,bm}
\graphicspath{{converted_graphics/}}
\usepackage{graphicx}
\begin{document}

\maketitle

\begin{abstract}
The dynamics of a flexible filament sedimenting in a viscous fluid are explored analytically and numerically. Compared to the well-studied case of sedimenting rigid rods, the introduction of filament compliance is shown to cause a significant alteration in the long-time sedimentation orientation and filament geometry. A model is developed by balancing viscous, elastic, and gravitational forces in a slender-body theory for zero-Reynolds-number flows, and the filament dynamics are characterized by a dimensionless elasto-gravitation number. Filaments of both non-uniform and uniform cross-sectional thickness are considered. In the weakly flexible regime, a multiple-scale asymptotic expansion is used to obtain expressions for filament translations, rotations, and shapes. These are shown to match excellently with full numerical simulations. Furthermore, we show that trajectories of sedimenting flexible filaments, unlike their rigid counterparts, are restricted to a cloud whose envelope is determined by the elasto-gravitation number. In the highly flexible regime we show that a filament sedimenting along its long axis is susceptible to a buckling instability. A linear stability analysis provides a dispersion relation, illustrating clearly the competing effects of the compressive stress and the restoring elastic force in the buckling process. The instability travels as a wave along the filament opposite the direction of gravity as it grows and the predicted growth rates are shown to compare favorably with numerical simulations. The linear eigenmodes of the governing equation are also studied, which agree well with the finite-amplitude buckled shapes arising in simulations.
\end{abstract}

\begin{keywords}
sedimentation; slender-body theory; Euler-Bernoulli elasticity; buckling
\end{keywords}

\section{Introduction}
The deformation and transport of elastic filaments in viscous fluids play central roles in many biological and technological processes. In cellular biology, stiff biopolymers such as actin and microtubules confer to cells their mechanical properties \citep{Gardel05} and are essential for functions as diverse as cell division, differentiation and morphogenesis \citep{Reinsch98,Shinar11}, cell motility \citep{Brennen77,Lauga09}, reproduction \citep{Fauci06,Gaffney11}, mucus transport \citep{Fulford86}, wound healing \citep{Ehrlich77}, and hearing \citep{Tilney92}, among others. In engineering applications, solutions of flexible and semiflexible polymers are commonly used for their non-Newtonian rheological properties \citep{Bird87}, which can lead to a variety of complex flow behaviors including hydrodynamic instabilities \citep{Shaqfeh96,Pan13} and chaotic mixing \citep{Groisman00,Thomases11}.

Of particular interest to us in this work are slender elastic filaments that are both compliant and inextensible: such is the case, to a first approximation, of stiff biological polymers such as actin and microtubules, and of a wide range of polymers used in engineering including xantham gum and carbon nanotubes. When such filaments are placed in a fluid flow or external field, the competition of external forces, viscous stresses, and internal elastic forces can result in complex deformations and dynamics, which in turn can have a significant impact on the macroscopic transport properties of large-scale suspensions. There have been many studies, both experimental and theoretical, of the dynamics of such filaments in various types of microscale flows, including simple shear flow \citep{Hinch76,bs01,ts04,Munk2006,Young09,Harasim13}, extensional flows \citep{KantslerGoldsteinFilaments,gkss12}, pressure-driven channel flows \citep{Steinhauser12}, vortex arrays \citep{Young07,Wandersman2010,Manikantan13}, and other more complex microfluidic flows \citep{Autrusson11,Wexler13}. Others have considered the case of a filament subject to either external or internal forces, such as forcing of various types at the filament ends \citep{TensionStraighteningNelson,Wiggins98}, internal actuation \citep{Lauga07,sl10,Adhikari12}, two-body interactions \citep{lpll07}, and self-attraction as a result of capillary interactions \citep{esbl13}, to name a few.

Though seemingly simple, the sedimentation of elastic filaments in a constant and uniform gravitational field has received limited attention and has yet to be fully analyzed even in the case of isolated filaments. The sedimentation of rigid fibers has been the subject of many studies and is well understood. At zero Reynolds number a rigid fiber with unit director $\hat{\boldsymbol{t}}$ sedimenting under gravity in an unbounded fluid will maintain its orientation and travel at a constant velocity $\v{U}=[\mu_{\perp}(\t{I}-\hat{\v{t}}\hat{\v{t}})+\mu_{\parallel}\hat{\v{t}}\hat{\v{t}}]\bcdot\v{F}_{G}$, where $\v{F}_{G}$ is the net gravitational force on the particle. The mobility coefficients $\mu_{\perp}$ and $\mu_{\parallel}$ depend on the exact shape of the particle ($\mu_{\parallel}\approx 2\mu_{\perp}$ for a slender body) \citep{Happel65}. Because of its drag anisotropy, the particle does not generally translate in the direction of gravity, but rather at a fixed angle $\theta$ that depends on its orientation ($\theta=0$ when the fiber is either parallel or perpendicular to gravity). As first predicted by \cite{Koch89}, this very simple picture is seriously complicated when multiple rigid fibers are allowed to interact hydrodynamically. In that case, long-range interactions drive a concentration instability as a result of the coupling between the orientation of the particles, which determines their settling direction, and the disturbance flows they drive in the fluid, which reorient them. This instability is indeed observed in both experiments \citep{Metzger05} and simulations \citep{Saintillan06,gt09} and takes the form of dense dynamic particle clusters which settle at significantly higher speeds than isolated particles. Even two sedimenting particles can undergo complex periodic sedimentation dynamics \citep{jspst06}.

Any small amount of flexibility is expected to qualitatively change the dynamics described above, even for an isolated filament. If the filament is allowed to bend as it sediments, this loss of symmetry will result in a coupling between its translational and rotational motions, leading to reorientation of the filament with respect to the direction of gravity. Because the orientation of the filament directly determines the direction of its velocity, we can also expect a non-trivial translational motion in both vertical and horizontal directions. The effect of fiber asymmetry on the dynamics was demonstrated experimentally by \cite{tsvk11} using rigid curved fibers, where particle rotations and unsteady trajectories were reported; complex spatial dynamics of curved fibers have also been observed in other situations such as in simple shear flow \citep{Wang12}. The situation is yet more complex in the presence of flexibility, as the filament shape and  grand mobility matrix evolve dynamically in time. Using a model based on the slender-body theory of \cite{Cox70}, \cite{xn94} argued that this coupling should cause a weakly flexible filament to reorient in a direction perpendicular to gravity regardless of its initial configuration and to assume a steady U-shape that depends on the relative magnitude of gravitational and elastic forces. These predictions were confirmed by \cite{lpl05} and \cite{Schlagberger05} using numerical simulations based on a discrete model of a filament as a string of rigidly connected beads with bending moments. However, a complete theoretical description of the shape evolution and reorientation dynamics and their influence on spatial trajectories has yet to be realized.

The case of a floppy filament with weak bending resistance is even more challenging, as large deformations may occur. For reasons that will be made clear in \S 5, a filament oriented parallel to gravity is subject to a compressive tension profile which, in some cases, may overcome bending resistance and lead to a buckling instability, much like that for a macroscopic Euler beam \citep{Love92}. Buckling of elastic filaments in viscous fluid flows has already been reported in a number of situations. \cite{bs01} simulated the dynamics of isolated elastic filaments in simple shear flow and showed that buckling occurs when the filaments are aligned with the axis of compression of the flow, resulting in normal stress differences; these dynamics were also shown to persist at finite concentration  \citep{ts04}. A theoretical analysis of this buckling was later provided by \cite{Young07} in a simpler setting, namely at the hyperbolic stagnation point of a two-dimensional linear extensional flow when the filament is initially aligned with the axis of compression. They demonstrated that above a critical flow strength compressive viscous forces indeed induce buckling, and showed that a series of unstable modes characterized by increasingly higher wavenumbers can become excited as the strain rate is increased. Instability was also observed in more complex flows such as vortex arrays in both simulations \citep{Young07,Manikantan13} and experiments \citep{Wandersman2010}, where it was shown to have a strong impact on the spatial transport of the filaments. As we discuss in \S 5, a similar buckling instability is also predicted under sedimentation for nearly vertical floppy filaments, though more complex asymmetric mode shapes are expected as the base tension profile can be shown to be compressive only over  the leading half of the filament.

The paper is organized as follows. In \S 2 we describe the energetics of a single flexible filament under the influence of gravity and derive the equations for the filament position and tension. The dynamics of the filament are characterized by a dimensionless quantity which we term the elasto-gravitation number. Filaments of both non-uniform and uniform cross-sectional thickness are considered. The numerical method used to solve for the filament shapes and dynamics is the topic of \S 3. In \S 4 we study weakly flexible filaments, where the elasto-gravitation number is large, and show that the introduction of filament compliance can alter dramatically the long-time sedimentation orientation and velocity. Equilibrium shapes are derived, and the assumption of timescale separation allows for predictions of slowly varying filament shapes and rotation rates. The buckling instability of a sedimenting filament is studied in \S 5, where a linear stability analysis is used to predict the most unstable waveforms, growth rates, and wave speeds; the results are shown to compare favorably with numerical simulations. We conclude with a discussion in \S 6.

\section{Mathematical formulation}
To model the dynamics of a slender elastic filament in a viscous fluid, we first describe the energetics of the system from which a local force balance may be derived. We then proceed to discuss the model for the fluid-body interactions; namely, we solve the Stokes equations of viscous flow using the slender-body theory of \cite{Johnson80}.

\subsection{Energy functional and local force balance}

Consider a filament of length $L$ with a centerline described by $\v{x}(s,t)$, where $s$ is the arc length and $t$ is time. The filament is assumed to be radially symmetric at each cross-section with a thickness given by $a\cdot r(s)$ (with $r(s)$ dimensionless). The following functional describes the energetics of the system,
\begin{gather}
\begin{split}
\mathcal{E}= &\, \frac{1}{2}\int_0^L B(s)|\v{x}_{ss}|^2\,\mathrm{d}s+\frac{1}{2}\int_0^L T(s)(|\v{x}_s|^2-1)\,\mathrm{d}s \\  & -\int_0^L \v{f}(s)\boldsymbol{\cdot} \v{x}(s)\,\mathrm{d}s-\int_0^L \v{F}_g(s) \boldsymbol{\cdot} \v{x}(s) \,\mathrm{d}s,
\end{split} \label{Energy}
\end{gather}
where index $s$ denotes differentiation with respect to arc length. The first term corresponds to a Hookean bending energy, proportional to the curvature of the filament: $B(s)=EI(s)$ is the bending stiffness, with $E$ the elastic modulus and $I(s)=\upi a^4 r(s)^4/4$ the area moment of inertia. The second term imposes filament inextensibility, with the tension $T(s)$ acting as a Lagrange multiplier. The third term is due to the fluid force per unit length $\v{f}(s)$ acting on the body at station $s$. Finally, the last term is a gravitational potential energy, where $\v{F}_g(s)=- \upi a^2 r(s)^2  \Delta \rho \,g\, \v{\hat{y}}$. Here $\Delta \rho$ is the density difference between the filament and the fluid, and $g>0$ is the gravitational acceleration.

By the principle of virtual work, the pointwise force on the filament is found by taking a variational derivative of the energy \eqref{Energy}. Perturbing $\v{x}$ by $\varepsilon \v{h}(s)$ and taking $\varepsilon\rightarrow 0$, we find
\begin{align}
\begin{split}
\frac{\delta \mathcal{E}}{\delta \v{x}}=&\int_0^L B(s)\v{x}_{ss}\bcdot \v{h}_{ss}\,\mathrm{d}s+\int_0^L T(s)\v{x}_s\bcdot \v{h}_s\,\mathrm{d}s-\int_0^L (\v{f}(s)+\v{F}_g(s))\bcdot \v{h}\,\mathrm{d}s\\
=&\int_0^L \Big[-(T(s)\v{x}_{s})_s+(B(s)\v{x}_{ss})_{ss}-\v{f}(s)-\v{F}_g(s)\Big]\bcdot \v{h}\,\mathrm{d}s\\
&+\Big[B(s)\v{x}_{ss}\bcdot \v{h}_s +\left(T(s)\v{x}_s-(B(s)\v{x}_{ss})_s\right)\bcdot \v{h}\Big]_0^L.
\end{split}
\end{align}
Setting the above to zero for all perturbations $\v{h}(s)$, we see that the fluid force acting on the filament is given by
\begin{gather}
\v{f}(s)=-\v{F}_g(s)-(T(s) \v{x}_s)_s+(B(s)\v{x}_{ss})_{ss},
\end{gather}
and we also observe the boundary conditions for solvability,
\begin{gather}
(B\v{x}_{ss})(0)=0, \,\,\,\,(B\v{x}_{ss})(L)=0,\label{bcs1}\\
(T\v{x}_s)(0)=(B\v{x}_{ss})_s(0), \,\,\,\,\,(T\v{x}_s)(L)=(B\v{x}_{ss})_s(L).\label{bcs2}
\end{gather}
As expected, the integrated fluid force along the filament is equivalent to the net gravitational force,
\begin{gather}
\int_0^L \v{f}\,\mathrm{d}s=\int_0^L \Big[-\v{F}_g(s)-(T(s) \v{x}_s)_s+(B\v{x}_{ss})_{ss}\Big]\,\mathrm{d}s=-\int_0^L \v{F}_g(s) \,\mathrm{d}s=-\v{F}_G.
\end{gather}
Scaling lengths upon $L$ and forces upon the total gravitational force, $F_G=|\v{F}_G|$, the dimensionless fluid force per unit length on the filament is given by
\begin{gather}
\bar{\v{f}}(\bar{s})=-\bar{\v{F}}_g(\bar{s})-(\bar{T}(\bar{s}) \bar{\v{x}}_{\bar{s}})_{\bar{s}}+\beta (\bar{B}(\bar{s})\bar{\v{x}}_{\bar{s}\bar{s}})_{\bar{s}\bar{s}},\label{f}
\end{gather}
where $\v{x}=L\bar{\v{x}}$, $s=L \bar{s}$, $T=F_G \bar{T}$, $B=(\upi/4)E a^4 \bar{B}$, and $\bar{\v{F}}_g(\bar{s})$ integrates to $-\v{\hat{y}}$. Here we have introduced an elasto-gravitation number, $\beta=\upi E a^4/(4 F_GL^2)$, which compares the elastic forces acting on the filament to the gravitational force. With all variables now understood to be dimensionless, we drop the bars in \eqref{f} for the duration of the paper.

\subsection{Fluid-body interaction and filament dynamics}
As a filament settles in a fluid, the elasto-gravitation forces acting along the body are coupled to the body's orientation and shape dynamics. When the Reynolds number is small ($Re=\rho U L/\mu \ll 1$, with $U$ a characteristic speed and $\mu$ the fluid viscosity), the fluid flow is well-described by the Stokes equations,
\begin{gather}
-\boldsymbol{\nabla} p +\mu \Delta \v{u}=0,\ \ \boldsymbol{\nabla} \bcdot \v{u}=0,
\end{gather}
where $\v{u}$ is the fluid velocity and $p$ is the pressure. We assume the filament moves in an infinite quiescent fluid, and the boundary conditions are the no-slip condition on the filament surface and $\v{u}(\v{x})\rightarrow \v{0}$ as $|\v{x}|\rightarrow \infty$. Classical works have developed slender-body theories for the velocities of slender filaments and the associated viscous forces along the filament length \citep{Cox70,Batchelor70,kr76,Johnson80}. More recently, \cite{ts04} coupled the dynamics of a flexible filament with the slender-body theory of viscous fluid-body interactions in an environment absent of gravity. Using the small aspect ratio of the filament as a small parameter, these asymptotic theories result in a relationship between the velocity of the filament centerline and the viscous force along the entire body length through a one-dimensional integral equation.

Scaling time upon a sedimentation timescale of $8 \upi \mu L^2/F_G$, the dimensionless velocity of a point $s$ along the body centerline is approximated as
\begin{gather}
\v{x}_t= -\t{\Lambda}[\v{f}]-\t{K}[\v{f}], \label{x_t}
\end{gather}
where $\v{f}$ is the scaled fluid force acting on the body given by \eqref{f} \citep{Johnson80}. This expression is accurate to order $O(\e^2)$ for the force $\v{f}$ and $O(\e^2\log(\e))$ for the velocity $\v{x}_t$, where $\e=a/L \ll 1$ is the body aspect ratio. The local and nonlocal operators in \eqref{x_t} are given by
\begin{gather}
\t{\Lambda}[\v{f}](s)=\left[(c(s)+1)\t{I}+(c(s)-3)\v{\hat{s}}(s)\v{\hat{s}}(s)\right]\bcdot \v{f}(s),\label{lambda[f]}\\
\t{K}[\v{f}](s)=\int_0^1 \left(\frac{\t{I}+\v{\hat{R}}(s,s')\v{\hat{R}}(s,s')}{|\v{R}(s,s')|}\bcdot\v{f}(s')-\frac{\t{I}+\v{\hat{s}}(s)\v{\hat{s}}(s)}{|s-s'|}\bcdot\v{f}(s)\right)\,\mathrm{d}s',\label{K[f]}
\end{gather}
where $\v{\hat{s}}=\v{x}_s$, $\v{R}(s,s')=\v{x}(s)-\v{x}(s')$, $\hat{\v{R}}(s,s')=\v{R}(s,s')/|\v{R}(s,s')|$, $c(s)=\log(4 s(1-s)/\e^2r(s)^2)$, and $\v{\hat{s}}\v{\hat{s}}$ and $\v{\hat{R}}\v{\hat{R}}$ are dyadic products. Using the local inextensibility condition $\v{x}_s\cdot \v{x}_s=1$, the filament position equation \eqref{x_t} can be manipulated to give an equation for the tension,
\begin{align}
\begin{split}
-2(&c-1) T_{ss}+(c+1)|\v{x}_{ss}|^2 T-2 c_s T_s- \v{x}_s\bcdot \partial_s\t{K}[(T \v{x}_s)_s]\\
= &\,(7c-5)\beta B(s)\v{x}_{ss}\bcdot \v{x}_{ssss}+6 (c-1)\beta B(s)|\v{x}_{sss}|^2+6\beta c_s B(s)\v{x}_{ss}\bcdot \v{x}_{sss}\\
&+\beta(4c_s B_s+(5c-3)B_{ss})|\v{x}_{ss}|^2+4(4c-3)\beta B_s \v{x}_{ss}\bcdot \v{x}_{sss} -\beta \v{x}_s\bcdot \partial_s\t{K}[(B\v{x}_{ss})_{ss}]\\
&+(c-3)\v{x}_{ss}\bcdot \v{F}_g+2(c-1) \v{x}_{s}\bcdot \partial_s\v{F}_g+2 c_s\v{x}_s\bcdot \v{F}_g+\v{x}_s\bcdot \partial_s\t{K}[\v{F}_g(s)].\label{tension}
\end{split}
\end{align}

If the filament is cylindrical with constant cross-section ($r(s)=1$), then $\v{F}_g(s)=-\v{\hat{y}}$ and $B(s)=B$ are constants,
while $c(s)=\log(4 s(1-s)/\e^2)$ varies, though the slender-body theory loses accuracy at the endpoints in this
case \citep{Johnson80}. Instead, if the filament thickness is described by the spheroidal profile
$r(s)=2\sqrt{s(1-s)}$, we have $c(s)=c=\log(1/\e^2)$, a constant. For such a filament shape, assuming uniform
material distribution, the gravitational force is spatially varying, $\v{F}_g(s)=-6 s(1-s)\v{\hat{y}}$, as is the bending stiffness, $B(s)=r(s)^4=16 s^2(1-s)^2$. In this case the boundary condition \eqref{bcs1} disappears. This limiting case is singular and is associated with an elastic boundary layer at the endpoints.

Finally, for convenience, we define here two integral operators that will appear in the asymptotic evaluation of \eqref{K[f]},
\begin{gather}
S[g](s)=\int_0^1\frac{g(s')-g(s)}{|s'-s|}\,\mathrm{d}s',\,\,\,\,\,\,\,\,P[g](s)=\int_0^1\frac{\Delta g(s,s')-g_s(s')}{|s'-s|}\,\mathrm{d}s',\label{S[]}
\end{gather}
where
\begin{gather}
\Delta g(s,s')=\frac{g(s)-g(s')}{s-s'} \cdot
\end{gather}

\section{Numerical method}

The governing equations are solved numerically using a variation of the method suggested by  \cite{ts04}. We denote by a superscript \textit{n} quantities at time $t_n$. Given the filament position at $t=t_n$, the tension $T^n(s)$ is first determined by solving a modification of equation~\eqref{tension},
\begin{align} \label{eq:tension_equation2}
\begin{split}
-2(c&-1) T^n_{ss}+(c+1)|\v{x}_{ss}|^2 T^n- \v{x}_s\bcdot \partial_s\t{K}_{\delta}[(T^n \v{x}_s)_s]\\
= &\,(7c-5)\beta\v{x}_{ss}\bcdot \v{x}_{ssss}+6 (c-1)\beta|\v{x}_{sss}|^2 -\beta\v{x}_s\bcdot \partial_s\t{K}_{\delta}[\v{x}_{ssss}]+(c-3)\v{x}_{ss}\bcdot \v{F}^n_g\\
&+2(c-1) \v{x}_{s}\bcdot \partial_s\v{F}^n_g+\v{x}_s\bcdot \partial_s\t{K}_{\delta}[\v{F}^n_g(s)] + \sigma(1-\v{x}_s\bcdot\v{x}_s),
\end{split}
\end{align}
where the position $\v{x}$ and its derivatives are evaluated at time $t_n$. We have chosen $c(s)=c$ a constant and $\v{F}_g(s)=-6s(1-s)\v{\hat{y}}$ as previously described, but we have assumed $B(s)=1$, an approximation that we justify later. Furthermore, we have added a restoring spring force (with a fitted parameter $\sigma$) that acts to correct numerical errors to filament inextensibility. We have introduced a regularized integral operator $\t{K}_\delta[\v{f}]$, where
\begin{equation}\label{eq:reg_nonlocal}
\t{K}_{\delta}[\v{f}](s)=\int_0^1\left(\frac{\t{I}+\hat{\v{R}}(s,s')\hat{\v{R}}(s,s')}{\sqrt{|\v{R}(s,s')|^2+\delta^2}}\bcdot\v{f}(s')- \frac{\t{I}+\hat{\v{s}}(s)\hat{\v{s}}(s)}{\sqrt{|s-s'|^2+\delta^2}}\bcdot\v{f}(s)\right)\,\mathrm{d}s'.
\end{equation}
A constant regularization parameter is chosen, $\delta=2\epsilon$, where $\epsilon$ is the filament aspect ratio. The error introduced by this regularization is $O(\e^2\log(\e))$ in the interior of the filament and $O(\e)$ near the filament ends, which could be further improved by use of a nonuniform regularization parameter $\delta(s)$ as in the work of \cite{ts04}. Discretizing the arc length as $s_j=j/N, \ j=0,1,...,N$, equations \eqref{eq:tension_equation2} and \eqref{eq:reg_nonlocal} are recast into a system of coupled linear equations for $T^n(s_j)$. We note that this system is dense owing to the nonlocal nature of the hydrodynamic interactions.

Following the solution of the tension at time $t_n$, the position of the filament at a time $t_{n+1}$ is then determined by a semi-implicit integration of equation~\eqref{x_t}. The stiffest part of the equation (the fourth derivative of the position) is treated implicitly, while the remaining terms such as the tension and lower derivatives of the position are extrapolated from previous data. The position at $t_{n+1}$ is given to second-order accuracy by
\begin{gather}\label{eq:time_marching}
\frac{1}{2 \Delta t}(3 \v{x}^{n+1}-4\v{x}^n+\v{x}^{n-1}) = \v{M}(2\v{x}^n-\v{x}^{n-1},\v{x}^{n+1}_{ssss}) + 2\v{N}(\v{x}^n)-\v{N}(\v{x}^{n-1}).
\end{gather}
Here, $\v{M}$ denotes all the terms that include the stiff operator, while $\v{N}$ collects the contributions of tension and gravity. The nonlocal integrals are evaluated explicitly at time $t_n$ and are supplied to equation~\eqref{eq:time_marching}. Solution of that equation then only requires inversion of a matrix of the form $-(1+c)\v{D}^4+(3-c)(2\v{x}_s^n-\v{x}_s^{n-1})(2\v{x}_s^n-\v{x}_s^{n-1})$, where $\v{D}$ is a finite difference operator. An Euler scheme is used for the first time step.

The spatial derivatives in equations \eqref{eq:tension_equation2} and \eqref{eq:time_marching} are discretized using second-order divided differences. Boundary conditions from \eqref{bcs1} and \eqref{bcs2} are translated onto the discrete points via one-sided finite differences. For all the results presented in this work, we use $N=256$, $\sigma=400$ and the dimensionless time step is $\Delta t = 10^{-5}$. For the linear stability results presented in \S 5, the time step is further reduced to $\Delta t = 10^{-6}$. The filament aspect ratio $\epsilon$ is fixed at $0.01$.

\section{Weakly flexible filaments}

It is a well known result that straight, rigid rods sediment in an infinite viscous fluid without any body reorientation \citep{kk91}. Curved filaments, however, have been shown to rotate during sedimentation until an equilibrium orientation is achieved \citep{tsvk11}. The introduction of filament flexibility, then, can result in filament shape changes but can also lead to complex body reorientation. To investigate the first effects of elasticity, we focus on the dynamics of a weakly flexible filament, where the elasto-gravitation number $\beta$ is assumed to be large.

\begin{figure}
\centering
$\begin{array}{cc}
\includegraphics[width=5in]{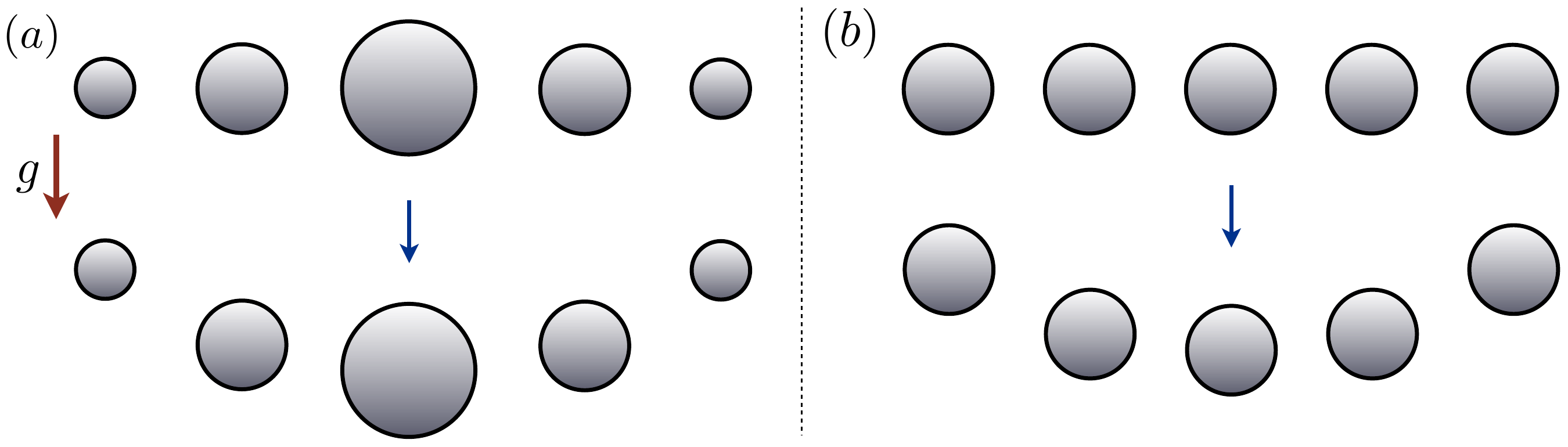}
\end{array}$
\caption{(Colour online) Illustration of the source of bending in model ``filaments.'' (\textit{a}) The leading-order effect: larger bodies sediment faster than smaller bodies in a viscous fluid, and filaments of non-uniform thickness will bend as a consequence. (\textit{b}) The secondary effect: the central bodies in a line of identical sedimenting spheres experience a stronger disturbance fluid flow, and will sediment faster than those near the ends.}
\label{Figure1}
\end{figure}

There are two different effects that can lead to shape changes of a weakly flexible sedimenting filament. As we will show, the leading-order effect is due to non-uniformity of the filament thickness along its length. Consider a filament of non-uniform thickness sedimenting in the direction of its minor axis, and for the sake of intuition consider as a simple model the distribution of sedimenting spheres of varying radii shown in figure~\ref{Figure1}(\textit{a}). In a highly viscous fluid, a sphere of radius $a$ settles with speed $U=2\Delta \rho g a^2/\mu$ (see \cite{Happel65}). The spheres near the center of the row will sediment faster than those near the ends, resulting in bending of the assemblage as depicted. A filament of uniform thickness is also expected to bend but as a consequence of a secondary effect, namely by nonlocal hydrodynamic interactions. Modeling such a filament as a row of \textit{identical} spheres, as illustrated in figure~\ref{Figure1}(\textit{b}), note that the disturbance flow experienced by the central spheres, due to the motion of the other spheres, will increase the sedimentation speed of the former. Bending from nonlocal hydrodynamics will be shown to be a higher-order effect. In this section we will study the behavior of spheroidal filaments, where $r(s)=2\sqrt{s(1-s)}$, while similar calculations for the case $r(s)=1$ are included in Appendix B.

\begin{figure}
\centering
$\begin{array}{cc}
\includegraphics{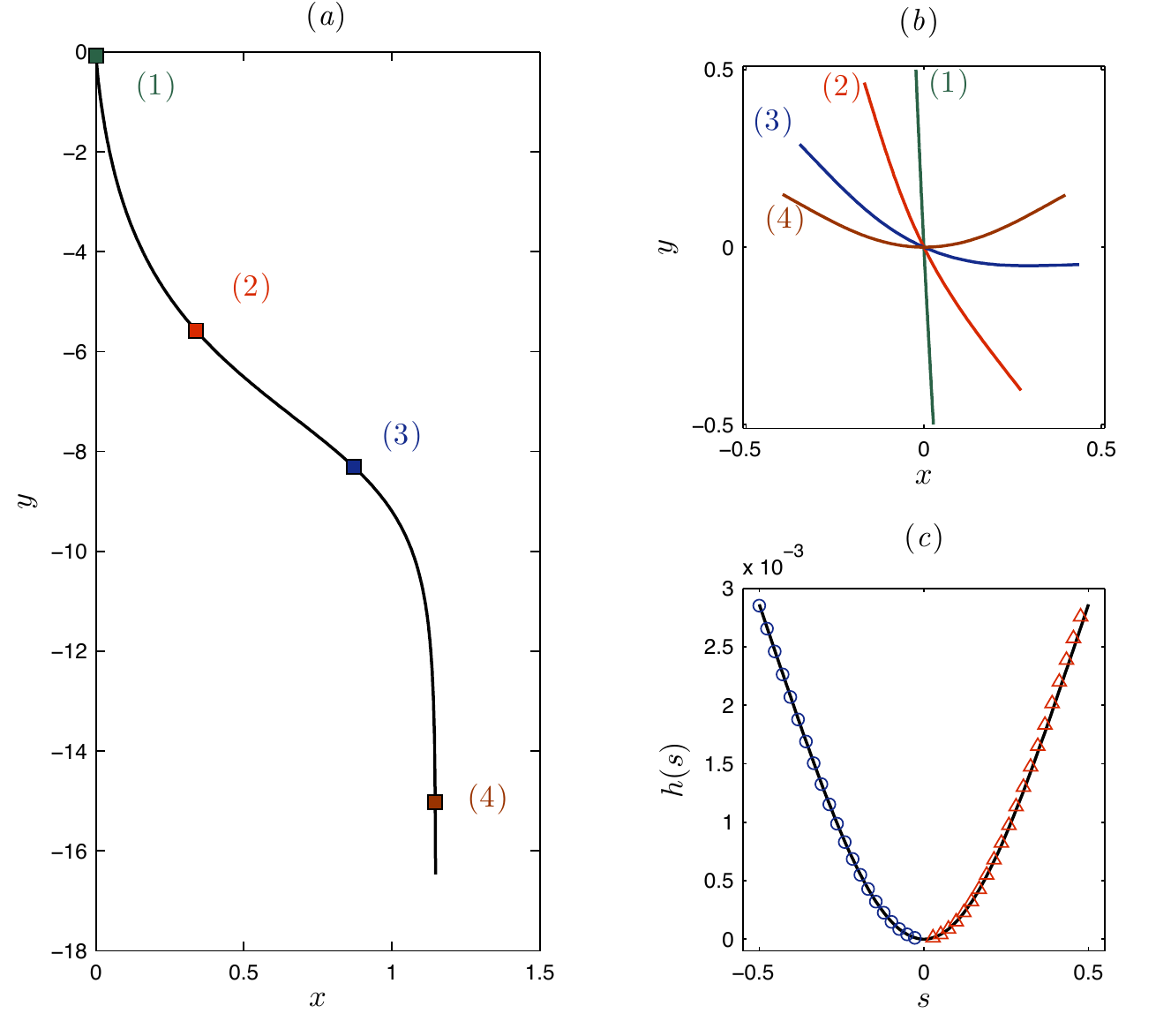}
\end{array}$
\caption{(Colour online) (\textit{a}) Trajectory of a filament with initial orientation angle $\theta_0=\upi/64$, with ``large'' elasto-gravitation number $\beta=0.02$, from numerical simulations. (\textit{b}) Shape of the filament as it sediments in a frame moving with its midpoint, also showing the reorientation process. Snapshots correspond to indicated points on the trajectory in (\textit{a}) (see also supplementary movie 1, which shows simultaneously the changing filament shapes and trajectories for two values of $\beta$, available at journals.cambridge.org/ßm). (\textit{c}) The final sedimenting shape, normalized by $\beta$, corresponding to $\beta=8\, ({\color{blue}\circ})$ and $\beta=0.02 \,({\color{red}\triangle})$ from simulations, along with the prediction from equation~\eqref{hB1} (solid line) showing the validity of the theory down to relatively small values of $\beta$.}
\label{Figure2}
\end{figure}

Returning to the full model described in \S 2, the complex interactions between shape changes and body reorientation can be seen in the numerical results of figure~\ref{Figure2}. In figure \ref{Figure2}(\textit{a}), an initially straight filament is released at the origin in a nearly vertical orientation and is allowed to deform and sediment freely under gravity. The initial angle between the tangent at the particle center and gravity is $\theta_0=\upi/64$, and we choose what we will find to be a relatively large value of $\beta=0.02$. As a result of its flexibility, weak deformations arise which cause the slow reorientation of the filament to a direction perpendicular to gravity, as shown in figure~\ref{Figure2}(\textit{b}). As the filament rotates away from its initial orientation, its settling motion incurs a lateral drift, which is strongest when the mean orientation forms an angle of approximately $\upi/4$ with the direction of gravity. As the filament eventually aligns horizontally, the drift slows and the trajectory asymptotes to a vertical line. Movie 1 in the supplementary materials shows the changing filament shapes and trajectories for two values of $\beta$, and is available online at journals.cambridge.org/ßm.

\begin{figure}
\centering
$\begin{array}{cc}
\includegraphics{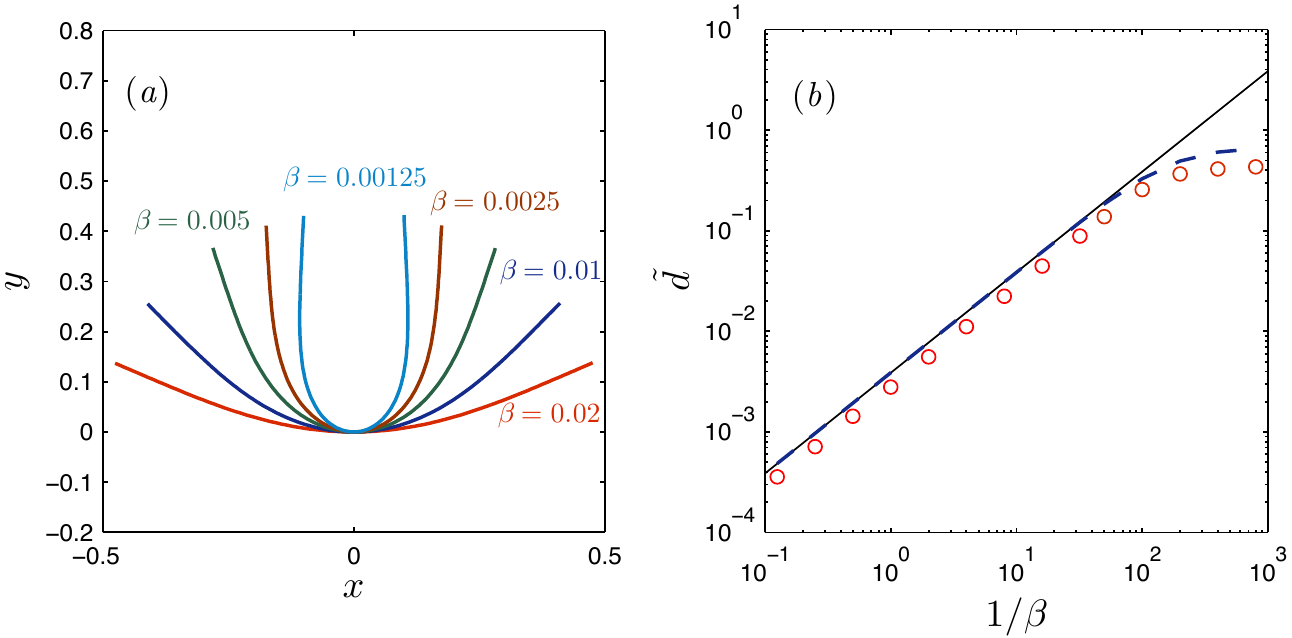}
\end{array}$
\caption{(Colour online) (\textit{a}) Steady-state shapes for $\beta$ in the range $0.00125 - 0.02$, obtained in numerical simulation with $r(s)=2\sqrt{s(1-s)}$ and $B(s)=1$. The deflections are shown in a frame moving with the filament midpoint. (\textit{b}) Maximum deflection of the filament shape $\tilde{d}=\max[\hat{\boldsymbol{n}}\bcdot\boldsymbol{d}(s)]$ as a function of $1/\beta$. Also shown by the solid line is the theoretical prediction following equation~(\ref{hB1}). The dashed line shows a correction to the theoretical prediction in which the filament shape was rescaled to preserve length.}
\label{Figure3}
\end{figure}

We observe in the weakly flexible regime that the only stable filament orientation is such that the body length is perpendicular to the direction of gravity. In this configuration, the filament assumes a symmetric, nearly parabolic shape as shown in figures~\ref{Figure2}(\textit{b})--(\textit{c}). Steady shapes in the limit of weak flexibility are found to collapse onto a single self-similar curve upon normalization by $\beta$ (as will be shown). The steady shapes of more flexible filaments are plotted in figure \ref{Figure3}(\textit{a}) for a decreasing sequence of values of the elasto-gravitation number, where more flexible filaments are seen to adopt horseshoe shapes. The final extent of bending can be characterized by the maximum deflection $\tilde{d}$ of the filament, which is shown in figure~\ref{Figure3}(\textit{b}) against $1/\beta$, exhibiting linear growth in the weakly flexible regime that extends as far down as $\beta \approx 0.02$. For elasto-gravitation numbers $\beta\lesssim 0.01$, the curve plateaus with the appearance of the horseshoe shape towards the maximum possible symmetric deflection value of one half. The weakly flexible regime may therefore be defined by ``large'' values of the elasto-gravitation number, $\beta \gtrsim 0.01$.

\subsection{Asymptotics in the weakly flexible regime: a separation of timescales}

We now set out to describe the filament shapes and dynamics analytically in the weakly flexible regime. As we have observed in the numerical simulations of figure~\ref{Figure2}, when the elasto-gravitation number is large the filament rotates in a time much longer than is required for the body to traverse many body lengths. Meanwhile, the filament is relatively stiff, so for a given orientation angle the body rapidly reaches its equilibrium shape. These observations suggest that there is a separation of timescales that will aid in the analysis of the system; the filament shape can be determined separately from the body rotation rate, and the rotation rate can be determined given a fixed body shape.

\begin{figure}
\centering
$\begin{array}{cc}
\includegraphics[width=2.4in]{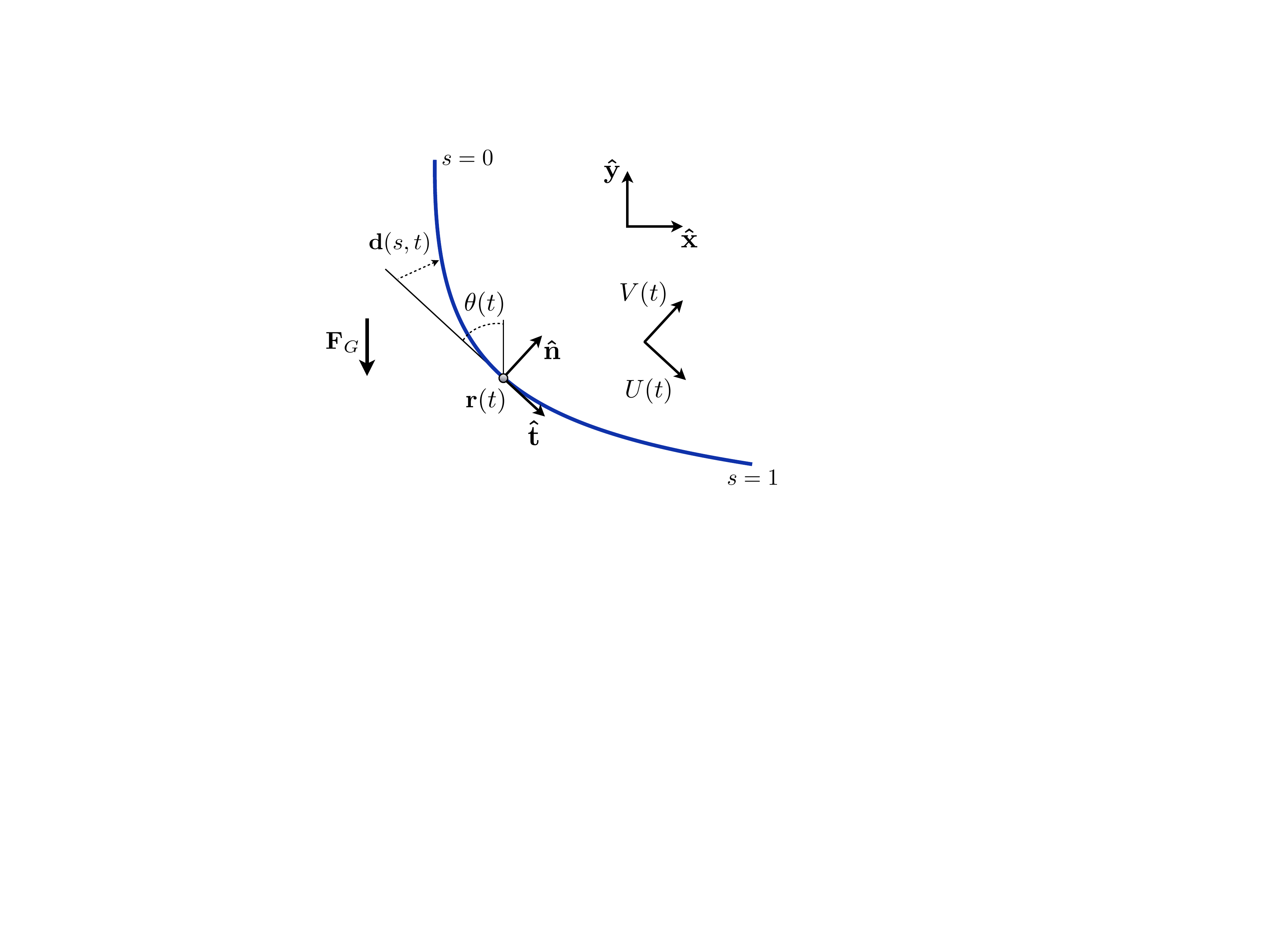}
\end{array}$
\caption{(Colour online) Filament illustration: the unit tangent and unit normal vectors at the single point $\v{r}(t)=\v{x}(s=1/2,t)$ are given by $\v{\hat{t}}$ and $\v{\hat{n}}$, respectively. $\theta(t)$ measures the angle between $\v{\hat{t}}$ and the direction of gravity, $-\v{\hat{y}}$.}
\label{Figure4}
\end{figure}

The position of the filament centerline at time $t$ can be written without loss of generality as
\begin{gather}
\v{x}(s,t)=\v{r}(t)+(s-1/2)\v{\hat{t}}(\theta(t))+\v{d}(s,t),\label{dstau}
\end{gather}
where $\v{r}(t)=\v{x}(1/2,t)$ is the position of the filament center, $\v{\hat{t}}=\v{x}_s(1/2, t)$ is the unit tangent vector there, and $\v{d}(s,t)$ is the time-dependent deviation of the filament from its straightened state (with $\v{d}(1/2,t)=0$). The filament is illustrated in figure~\ref{Figure4}. The natural coordinate system that rotates in time with the body is then described by
\begin{gather}
\v{\hat{t}}(\theta)=-\cos \theta\,\v{\hat{y}}+\sin\theta\,\v{\hat{x}},\\
\v{\hat{n}}(\theta)=\sin \theta\,\v{\hat{y}}+\cos\theta\,\v{\hat{x}},
\end{gather}
where $\theta=\theta(t)$ measures the angle between $-\v{\hat{y}}$ and the unit tangent vector $\v{\hat{t}}$, and $\v{\hat{n}}$ is the vector normal to the filament at its midpoint. The translational velocity of the midpoint is written as $\v{r}'(t)=\v{U}(t)=U(t)\v{\hat{t}}+V(t)\v{\hat{n}}$.

For large values of the elasto-gravitation number, $\beta \gg 1$, we observe three distinct timescales in the numerical simulations that motivate a multiple-scale analysis. The first is a very short elastic relaxation timescale of $O(\beta^{-1})$. The second is the timescale of $O(1)$ on which the body sediments a distance comparable with its length. The third is a very long timescale of $O(\beta)$ on which the body may reorient on account of its nontrivial shape. Our aim is to study the shape changes of the filament on the latter two timescales, during which the body translates and rotates through the fluid. Defining the scaled time $\tau=\beta^{-1}t$, we analyze the system by the method of multiple scales wherein variables are assumed to have a separate explicit dependence upon both $t$ and $\tau$ (see \cite{bo1999}). A uniform solution to equations \eqref{x_t} and \eqref{tension} is then sought by assuming regular expansions of the tension and filament shape in powers of the small number $\beta^{-1}$ of the form
\begin{gather}
T(s,t,\tau; \beta)=T^{(0)}(s,t,\tau)+\beta^{-1} T^{(1)}(s,t,\tau)+O\left(\beta^{-2}\right).
\end{gather}
Upon inspection of equation $\eqref{f}$, we observe that the sedimentation and elastic effects are balanced when $\beta(B(s)\v{d}_{ss})_{ss}=O(1)$. Hence, the deflection of the filament due to gravitational effects is $O(\beta^{-1})$, and we write
\begin{gather}
 \v{d}(s,t,\tau;\beta)=\beta^{-1}u(s,t,\tau)\v{\hat{n}}(\theta)+\beta^{-2}u_1(s, t, \tau)\v{\hat{n}}(\theta)+\beta^{-2}v_1(s, t, \tau)\v{\hat{t}}(\theta)+O\left(\beta^{-3}\right).
\end{gather}
Here we have also used the filament inextensibility, which requires that the only filament deflections at first order in $\beta^{-1}$ are normal to $\v{\hat{t}}$. The definition of $\v{d}$ implies $\v{d}_s(1/2,t,\tau;\beta)=0$. The translational velocity and orientation angle are similarly expressed,
\begin{gather}
U(t,\tau;\beta)=U^{(0)}(t,\tau)+\beta^{-1}U^{(1)}(t,\tau)+O\left(\beta^{-2}\right),\\
V(t,\tau;\beta)=V^{(0)}(t,\tau)+\beta^{-1}V^{(1)}(t,\tau)+O\left(\beta^{-2}\right),\\
\theta(t,\tau; \beta)=\theta^{(0)}(t,\tau)+\beta^{-1} \theta^{(1)}(t,\tau)+O\left(\beta^{-2}\right).
\end{gather}
Inserting the expressions above into equation \eqref{x_t}, and dotting separately with either $\v{\hat{t}}(\theta)$ and $\v{\hat{n}}(\theta)$, we find the leading-order relations,
\begin{gather}
U^{(0)}=2(c-1)\left[T^{(0)}_s-F_g\cos\theta^{(0)}\right]+2S\left[T^{(0)}_s-F_g\cos\theta^{(0)}\right],\label{eq:leadingorder1}\\
V^{(0)}+(s-1/2)\theta^{(0)}_t=-(c+1)\left[(Bu_{ss})_{ss}-F_g\sin\theta^{(0)}\right]-S\left[(Bu_{ss})_{ss}-F_g\sin\theta^{(0)}\right]\label{eq:leadingorder2},
\end{gather}
where $S[\cdot]$ is the nonlocal hydrodynamic contribution for a straight filament defined in \eqref{S[]}. Recall that $B=B(s)$, $F_g=F_g(s)$, and $c=\log(1/\e^2)$. Denoting by  $\mathcal{L}_n(s)$ the $n^{\mathrm{th}}$ shifted Legendre polynomial (defined on $s\in[0,1]$), we have $S[\mathcal{L}_n(s)]=\lambda_n \mathcal{L}_n(s)$, with $\lambda_n=-2\sum_{i=1}^n (1/i)$. Hence, the equations above are made tractable by expressing variables in the Legendre polynomial basis (see \cite{Gotzthesis}). Using the orthogonality of the Legendre polynomials, we recover the leading-order sedimentation velocity,
\begin{gather}
\v{U}^{(0)}=2(c-1)\cos\theta^{(0)}\, \v{\hat{t}}(\theta^{(0)})-(c+1)\sin\theta^{(0)}\, \v{\hat{n}}(\theta^{(0)}),\label{U0}
\end{gather}
and in addition we find
\begin{gather}
T^{(0)}_s-F_g\cos\theta^{(0)}=\cos\theta^{(0)}, \label{eq:tension0} \\
(Bu_{ss})_{ss}-(F_g+1)\sin\theta^{(0)}=\frac{s-1/2}{1-c}\theta_t^{(0)}.\label{eq:shape0}
\end{gather}
The case of uniform filament thickness, with $F_g(s)=-1$, is considered in Appendix B. The leading-order effect illustrated in figure~\ref{Figure1} is studied now by inserting $F_g(s)=-6s(1-s)$, which results in the leading-order tension,
\begin{gather}
T^{(0)}=s(1-2s)(1-s)\cos\theta^{(0)}.
\end{gather}
Meanwhile, multiplying equation \eqref{eq:shape0} by $(s-1/2)$ and integrating, we find $\theta^{(0)}_t=0$. The filament therefore does not rotate on the timescale $t$, but may still rotate on the longer timescale, $\theta^{(0)}(t,\tau)=\theta^{(0)}(\tau)$. The leading-order deflection of the filament from its straightened state can now be determined from equations \eqref{eq:shape0} and \eqref{bcs1},
\begin{gather}
u(s,\tau)=h(s)\sin\theta^{(0)}(\tau),
\end{gather}
with
\begin{gather}
B(s)h_{ss}=\frac{1}{2}s^2(1-s)^2,\\
h(1/2)=h_s(1/2)=0.
\end{gather}
If the filament is composed of a uniform material, a corresponding bending stiffness $B(s)=r(s)^4=16 s^2(1-s)^2$ then results in the filament deflection profile
\begin{gather}
h(s)=\frac{1}{64}\left(s-\frac{1}{2}\right)^2.\label{h0}
\end{gather}
Surprisingly, the shape of the filament is symmetric about its midpoint at leading order for any orientation, and the scaling of the deflection with the orientation angle is given simply by $\sin \theta^{(0)}(\tau)$. In order to determine the orientation angle $\theta^{(0)}(\tau)$, we must look to higher order. At $O(\beta^{-1})$, equation \eqref{x_t} yields the expression
\begin{align}
\begin{split}
V^{(1)}+&(s-1/2)\left(\theta^{(0)}_\tau+\theta^{(1)}_t\right)=(c+1)\left[(T^{(0)}u_s)_s-(Bu_{1,ss})_{ss}+\theta^{(1)}\cos\theta^{(0)}F_g\right]\\
&+(c-3)u_s\cos\theta^{(0)}+S\left[(T^{(0)}u_s)_s-(Bu_{1,ss})_{ss}+\theta^{(1)}\cos\theta^{(0)}F_g\right]\\
&+\cos\theta^{(0)}S[u_s]+
\cos\theta^{(0)}P[u],\label{V1}
\end{split}
\end{align}
where the integral operator $P[\cdot]$ is defined in equation \eqref{S[]}. Multiplying \eqref{V1} by $(s-1/2)$ and integrating, we have
\begin{gather}
\theta^{(0)}_\tau+\theta^{(1)}_t=\frac{A}{2}\sin(2\theta^{(0)}),\\\label{th1th0}
A=12((c-1)I_1+(c-5)I_2+I_3),
\end{gather}
where
\begin{align}
&I_1=\int_0^1(1-6s+6s^2)h(s)\,\mathrm{d}s,\\
&I_2=\int_0^1(s-1/2)h_s(s)\,\mathrm{d}s,\\
&I_3=\int_0^1(s-1/2)P[h](s)\,\mathrm{d}s.
\end{align}
The secular behavior in the expansion is removed by taking $\theta^{(1)}=\theta^{(1)}(\tau)$, and we are left with an equation for the dynamics of $\theta^{(0)}$,
\begin{gather}
\theta^{(0)}_\tau=\frac{A}{2}\sin(2\theta^{(0)}).\label{theta_0eqn}
\end{gather}
Inserting the expression for $h(s)$ obtained in equation \eqref{h0}, we have $I_1=1/1920$ and $I_2=I_3=1/384$, so that $A=3 (c-7/2)/80=3 (\log(1/\e^2)-7/2)/80$. The constant $A$ is positive (and the result is physical) in the slender-body regime, or specifically when: $\e<\exp(-7/4)\approx 0.17$. We therefore have that the orientation angle $\theta=0$ is unstable, and that $\theta=\pm \upi/2$ are stable. The filament, on a timescale $O(\beta/c)$, will reorient so that its central tangent vector is perpendicular to gravity.

In the calculation above, if we were to take the bending stiffness to be constant along the centerline ($B(s)=1$), we instead find
\begin{gather}
h(s)=\frac{1}{64}\left[\left(s-\frac{1}{2}\right)^2-\frac{4}{3}\left(s-\frac{1}{2}\right)^4+\frac{16}{15}\left(s-\frac{1}{2}\right)^6\right], \label{hB1}
\end{gather}
which matches equation \eqref{h0} in the interior of the filament, but predictably leads to a slightly smaller filament deflection from the horizontal plane. The corresponding orientation dynamics are still given by equation \eqref{theta_0eqn}, but now we have $I_1=1/2520$, $I_2=1/560$, and $I_3=101/50400$, so that $A=11(c-369/110)/420$. We still find $A>0$ in a similar range of body aspect ratios, $\e \lesssim  0.19$. This calculation is not to be confused with that for a filament of uniform thickness, as described in Appendix B. However, the similarity between \eqref{h0} and \eqref{hB1} suggests that computing with the assumption $B(s)=1$ even for a spheroidal body, which avoids the computational issues related to an elastic boundary layer, is reasonable. We therefore choose $B(s)=1$ for our computations for the remainder of the paper (and in the previous section). 

The body shapes predicted by \eqref{hB1} are shown in figure~\ref{Figure2}(\textit{c}) as a solid line, from which we see excellent agreement with the results of the numerical simulations (shown as symbols) down to $\beta\approx 0.02$. The maximum deflection of the filament shape is shown in figure~\ref{Figure3}(\textit{b}), with the results from the full simulations shown as circles and from the prediction as a solid line, which provides a quantitative measure of the accuracy and breakdown of the simple theory. At the order of our consideration the filament is not inextensible, and as a consequence we observe a systematic overestimation of the numerical results. A simple improvement of the prediction is obtained by rescaling the shape to unit length, as shown by a dashed line in figure~\ref{Figure3}(\textit{b}).

For $\beta \lesssim 0.01$, the shapes are no longer self-similar and depart significantly from the expression in \eqref{hB1}. Viscous stresses associated with the gravitational forcing are now strong enough to overwhelm the elastic stiffness, and a horseshoe-like shape emerges as seen from numerical simulations in figure \ref{Figure3}. The two ends of the filament approach one another for smaller $\beta$, and for $\beta\lesssim 0.001$ the filament can overlap itself unless steric effects are taken into account.

\subsection{Filament trajectories and particle clouds}

\begin{figure}
\includegraphics[scale=1]{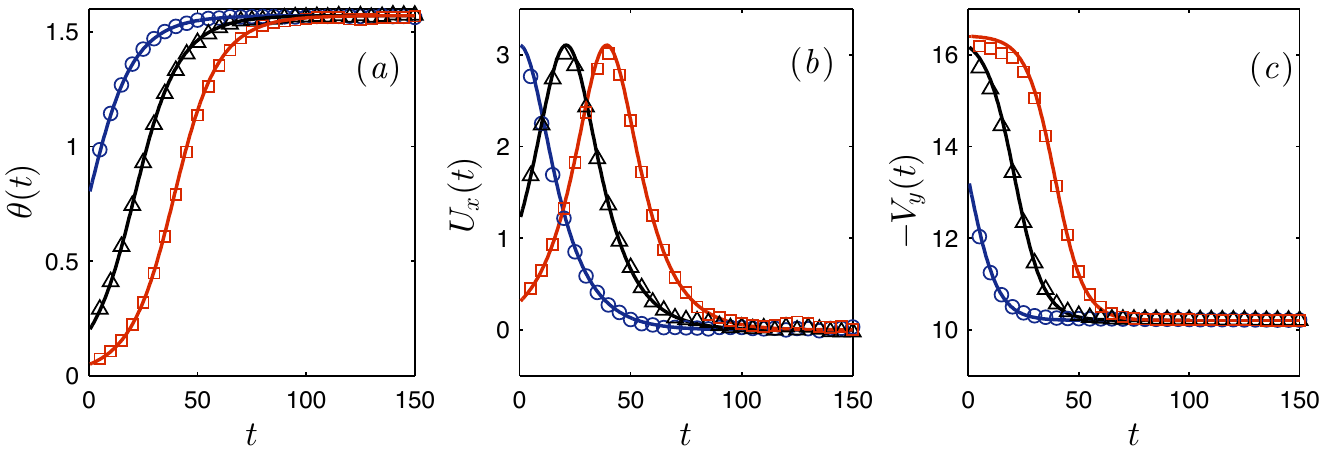}
\caption{Results from simulations at  $\beta=2$ and initial orientations $\upi/4\,({\color{blue}\circ})$, $\upi/16\,({\color{black}\triangle})$ and $\upi/64\,({\color{red}\square})$. Overlaid in solid lines are theoretical predictions for: (\textit{a}) filament orientation $\theta(t)$, (\textit{b}) horizontal filament velocity $U_x(t)$, and (\textit{c}) downward filament velocity $-V_y(t)$.}
\label{Figure5}
\end{figure}

We have shown that appreciable changes in the filament shape and orientation are found on the scale over which the filament sediments many body lengths through the fluid. Writing the dynamics only in terms of the single time $t$, the filament rotation rate at leading order is given by
\begin{gather}
\theta_t= \frac{A}{2\beta}\sin(2\theta),\label{theta_0eqn2}
\end{gather}
with $A=3 (c-7/2)/80$ for $B(s)=16 s^2(1-s)^2$. Integrating \eqref{theta_0eqn2} and setting $\theta(0)=\theta_0$, we find
\begin{gather}\label{eq:theta_t}
\tan(\theta(t))=\tan(\theta_0)\exp(At/\beta).
\end{gather}
The horizontal and vertical filament velocities were previously approximated to $O(1/\beta)$; inserting equation \eqref{theta_0eqn2} into \eqref{U0}, and writing $\v{U}(t)=U_x \v{\hat{x}}+V_y \v{\hat{y}}$, we find:
\begin{gather}
U_x(t)=\frac{(c-3)\tan(\theta_0)\exp(At/\beta)}{1+\tan^2(\theta_0)\exp(2A t/\beta)}\label{eq:Uvelocity},\\
V_y(t)=-(c+1)-(c-3)\left(\frac{1}{1+\tan^2(\theta_0)\exp(2A t/\beta)}\right)\label{eq:Vvelocity}.
\end{gather}
Integrating the velocities above leads to an approximation of the filament trajectory accurate to $O(1)$. Assuming that the filament is initially centered at the origin, the material point $s=1/2$ follows the path $(X(t),Y(t))$, where
\begin{gather}
\tan\left(\frac{A}{\beta(c-3)}X(t)+\theta_0\right)=\tan(\theta_0)\exp(At/\beta)\label{eq:X_traj},\\
Y(t)=-(c+1)t-\frac{\beta(c-3)}{2A}\log\left(\frac{(1+\tan^2(\theta_0))\exp(2 At/\beta)}{1+\tan^2(\theta_0)\exp(2 A t/\beta)}\right)\label{eq:Y_traj}.
\end{gather}
The full trajectory is described implicitly by the equation
\begin{align}
\begin{split}
\tan^{\alpha}(\theta_0)\sin(\theta_0)&\exp\left(-\frac{A}{\beta(c-3)}Y(t) \right)=\\
&\tan^{\alpha}\left(\frac{A}{\beta(c-3)} X(t)+\theta_0\right)\sin\left(\frac{A}{\beta(c-3)} X(t) +\theta_0\right),\label{trajectory}
\end{split}
\end{align}
where $\alpha=(c+1)/(c-3)>1$. In contrast to the constant horizontal velocity of a straight sedimenting rod, the filament drifts horizontally a finite distance (assuming $0 < \theta_0 \leq \upi/2$),
\begin{gather}
X(\infty)=\int_0^{\infty}U_x(t)\,\mathrm{d}t=\frac{\beta(c-3)}{A}\left(\frac{\upi}{2}-\theta_0\right),\label{X_infty}
\end{gather}
and $X(\infty)=0$ for $\theta_0=0$. The horizontal drift is monotonic in the initial orientation angle on this domain. The maximum drift is given for $\theta_0\rightarrow 0^+$, where $X(\infty)\rightarrow \upi \beta (c-1)/2A$. The drift is also monotonic in the elasto-gravitation number in this regime, with larger distances traversed by stiffer filaments, and $X(\infty)\rightarrow \infty$ for rigid fibers, $\beta\rightarrow \infty$.

\begin{figure}
\includegraphics{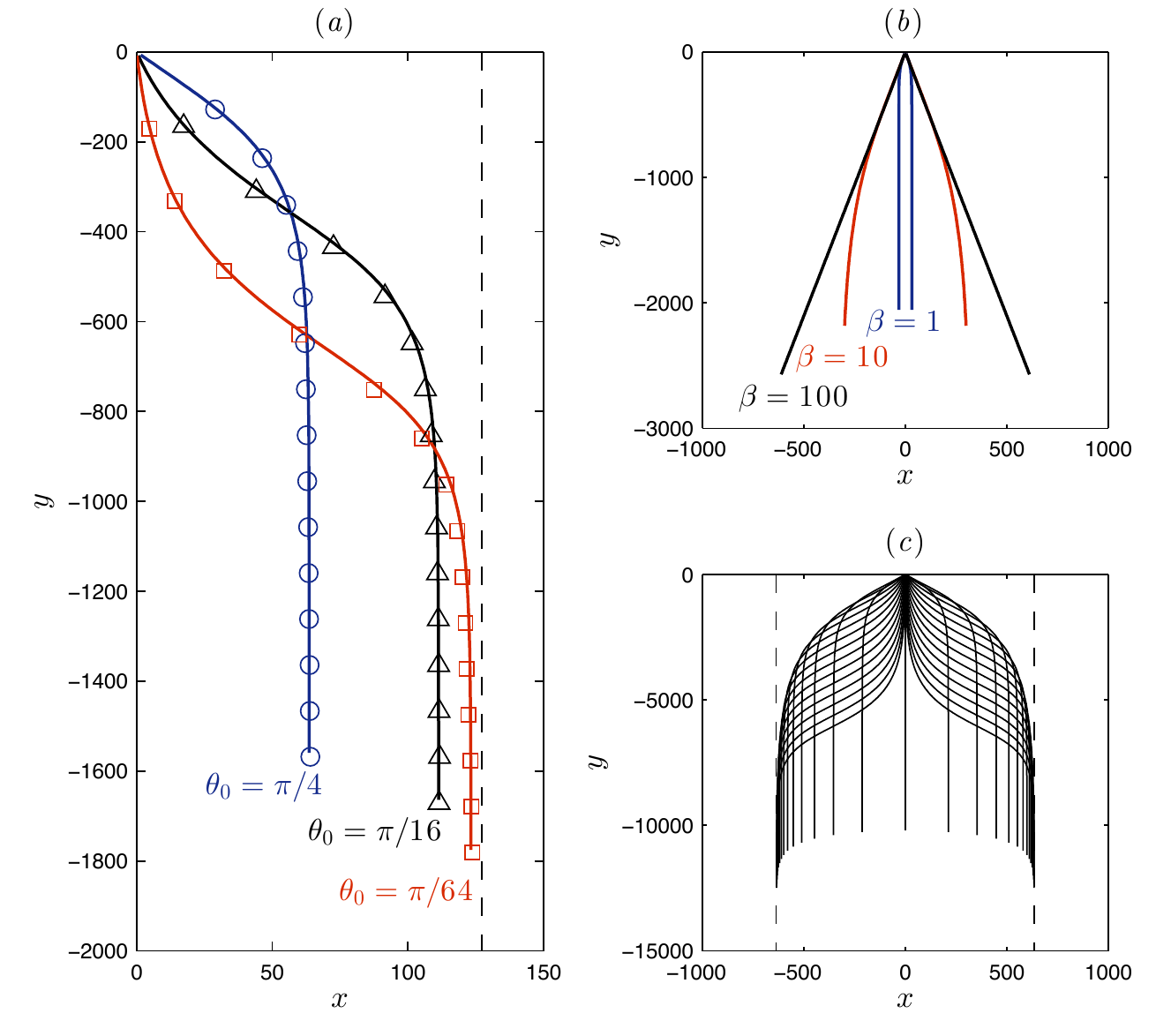}
\caption{(Colour online) (\textit{a}) Results from simulations (symbols) and from analytical predictions (solid lines) showing the trajectory of sedimentation of the midpoint of a filament for $\beta=2$ released with three different initial orientations.  The dashed line represents the maximum width of spreading as predicted by \eqref{X_infty}. (\textit{b}) Predicted trajectories for different values of $\beta$ when the filament is released at an angle of $\pm \upi/4$. (\textit{c}) A visualization of the predicted spreading of sedimenting flexible filaments. Here, $\beta=10$ and the initial angle of release varies in the range $[-\upi/2,\upi/2]$. Also shown is the maximum extent of cloud spreading. (Supplementary movie 2 shows the spreading of a cloud of weakly flexible non-interacting filaments with varying initial orientations.)}
\label{Figure6}
\end{figure}

The horizontal and vertical velocities and the filament rotation rate following equations \eqref{eq:theta_t}, \eqref{eq:Uvelocity} and \eqref{eq:Vvelocity}, respectively, are shown in figure~\ref{Figure5} for $\beta=2$ and three different initial orientations. Also shown are numerical results for these parameters that show excellent agreement with the predictions. One can clearly see from the figures that the filament initially drifts increasingly faster in a direction perpendicular to gravity, attaining a maximum horizontal velocity at approximately $\theta=\upi/4$. $U_x(t)$ then decreases to zero, which corresponds to the trajectory in figure~\ref{Figure2}(\textit{a}) tending asymptotically to a vertical line. The vertical velocity at this point settles to a constant value of $V_y(t\rightarrow \infty)=-(c+1)$, which gives $V_y\approx -10.21$ using $\epsilon=0.01$ as in the simulations. This value corresponds to the minimum speed of sedimentation in the entire process, corresponding to the drag being maximized in this regime for bodies sedimenting perpendicular to the long filament axis.

The monotonic increase of the span of spreading $X(\infty)$ with both the initial orientation and the elasto-gravitation number suggests interesting trajectories for filaments in this regime. Figure~\ref{Figure6} shows the trajectories associated with these dynamics. Numerical results for three different initial orientations, all for $\beta=2$, are shown in figure~\ref{Figure6}(\textit{a}) to match excellently with the predicted trajectories. Note again that the maximum width of spreading is attained for $\theta_0=0^{\pm}$, and the vertical asymptote of the trajectory approaches this value for small initial orientations. The qualitative difference between weakly flexible filaments and rigid rods is illustrated in figure~\ref{Figure6}(\textit{b}). With increasing values of the elasto-gravitation number, the trajectories of filaments placed at the same initial orientation ($\theta_0=\pm\upi/4$ in this case) approach the $\beta\rightarrow\infty$ limit of rigid rods, which sediment without rotating and at an angle that depends only on their initial orientation.

Finally, in figure~\ref{Figure6}(\textit{c}), we show how in this regime the lateral spreading of filament trajectories is confined to a cloud whose width is dictated by the elasto-gravitation number. The different trajectories correspond to different initial orientations with initially horizontal filaments sedimenting vertically downwards, and the widest spreading attained, as mentioned above, for $\theta_0=0^{\pm}$ (see movie 2 in the supplementary material, which shows the spread of weakly flexible filaments of varying initial orientation). Neglecting hydrodynamic interactions between bodies, consider the release of many filaments at the origin, with a probability density function of their orientations given by $\rho(\theta_0)$ on $\theta_0\in[0,\pi/2]$. Once the bodies have settled into their vertical trajectories, the radial distances from the origin (in the plane perpendicular to gravity) are distributed as $\rho(\theta_0) X(\infty)$. Assuming uniformly distributed filaments, $\rho(\theta_0)=\sin(\theta_0)$, then the radial distribution of the filament cloud as seen in figure~\ref{Figure6} is given by $[\beta(c-3)/A]\left(\frac{\upi}{2}-\theta_0\right)\sin(\theta_0)$. Integrating, the mean filament drift is given by $(\pi-2)\beta(c-3)/(2A)$, and the variance by $(\pi-3)\beta^2(c-3)^2/A^2$. 

\section{Buckling of flexible filaments}

\begin{figure}
\centering
\includegraphics[width=5.2in]{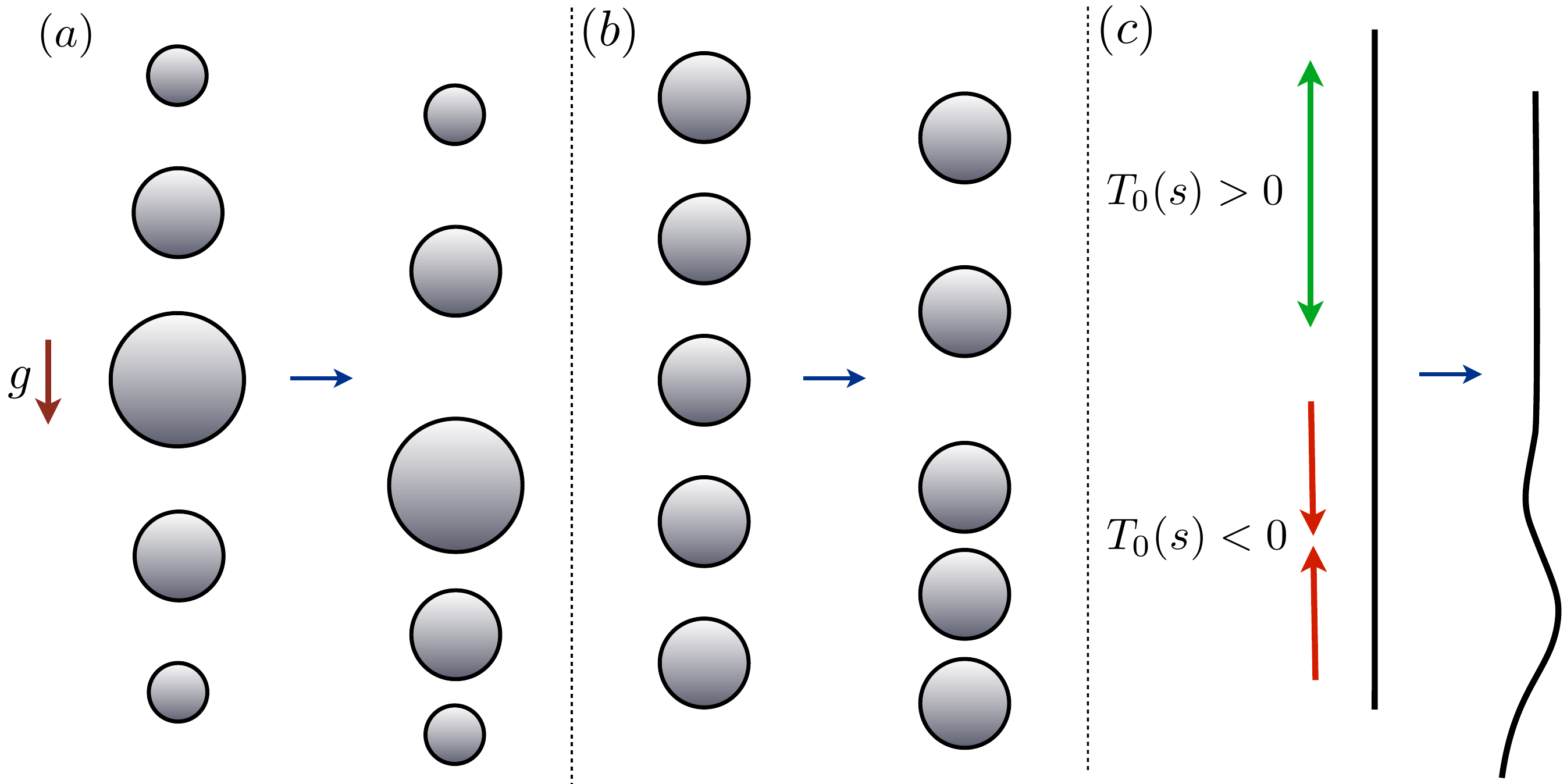}
\caption{(Colour online) Illustration of the source of tension and buckling in model ``filaments,'' as in figure~\ref{Figure1}. (\textit{a}) The leading-order effect: larger bodies sediment faster than smaller bodies in a viscous fluid. (\textit{b}) The secondary effect: the central bodies in a line of identical sedimenting spheres experience a stronger disturbance fluid flow, and will sediment faster than those near the ends. (\textit{c}) The effects in (\textit{a}) or (\textit{b}), along with inextensibility, can lead to buckling of a sedimenting filament.}
\label{Figure7}
\end{figure}

Our attention now turns to the opposite extreme, the case of extremely flexible filaments for which the elasto-gravitation number is small, $\beta \lll 1$. Of particular interest in this case is the possibility of a dramatic buckling event, which may be exhibited by an elastic body when compressive forces overcome its structural rigidity. Potential sources of a buckling instability in the context of sedimentation are illustrated in figure~\ref{Figure7}, and are identical to the sources of bending shown in figure~\ref{Figure1}. With spheres sedimenting according to their sizes, the array of spheres in figure~\ref{Figure7}(\textit{a}) will separate in the top half of the train, and collapse in the bottom half (in the direction of gravity). If the spheres are constrained so that their relative positions are fixed, there will be a positive tension in the top half of the train, and a negative (compressive) tension in the bottom half. This compression can cause a sufficiently flexible filament to buckle, as we show below. It is similarly argued that this source of instability will vanish if the filament density increases monotonically in the direction of gravity.

If the filament is of uniform thickness, the secondary effect from nonlocal hydrodynamic interactions can also lead to buckling. As illustrated in figure~\ref{Figure7}(\textit{b}), the spheres nearer to the center of the train sediment faster than those at the leading and trailing ends. This effect can also lead to buckling of a sufficiently flexible filament. Once again, the leading-order effect is now considered by studying a spheroidal filament shape, and comments on the case $r(s)=1$ are included in Appendix B.

\begin{figure}
\centering
\includegraphics{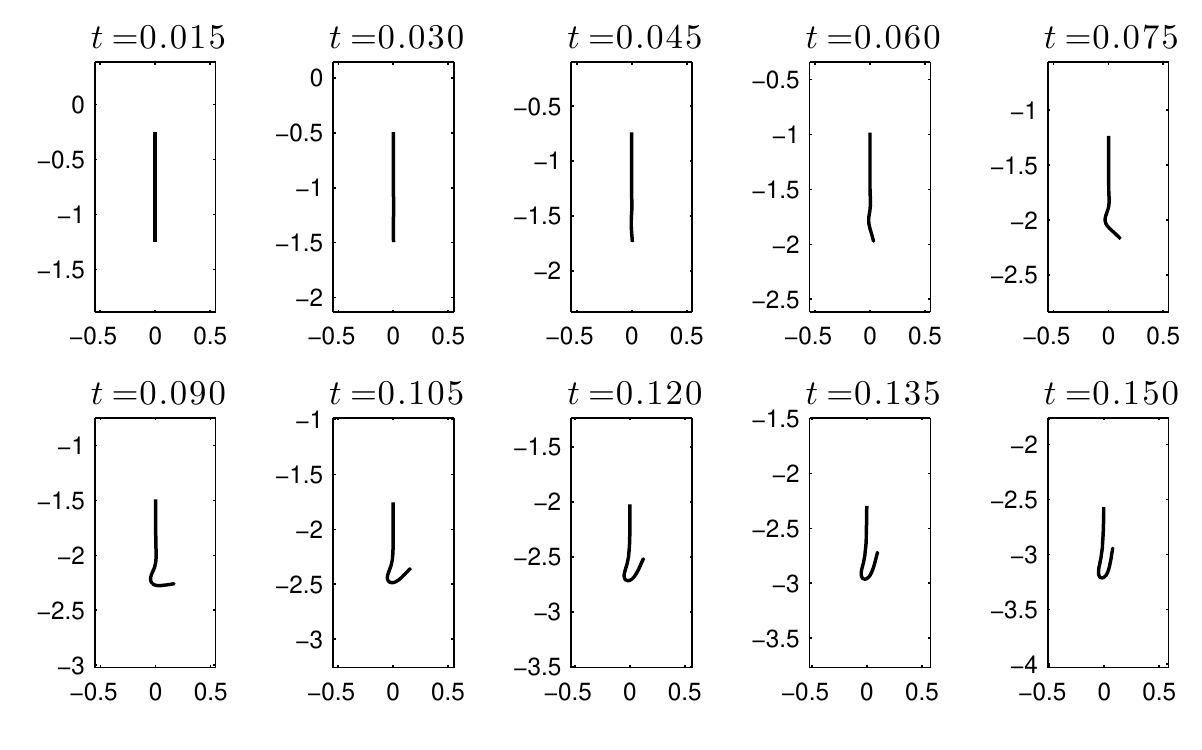}
\caption{Moderate buckling is observed in simulations for $\beta= 10^{-4}$ and $B(s)=1$. (See also supplementary movie 3, which shows the buckling of sedimenting filaments with three values of the elasto-gravitation number $\beta$.)}
\label{Figure8}
\end{figure}

\begin{figure}
\centering
\includegraphics{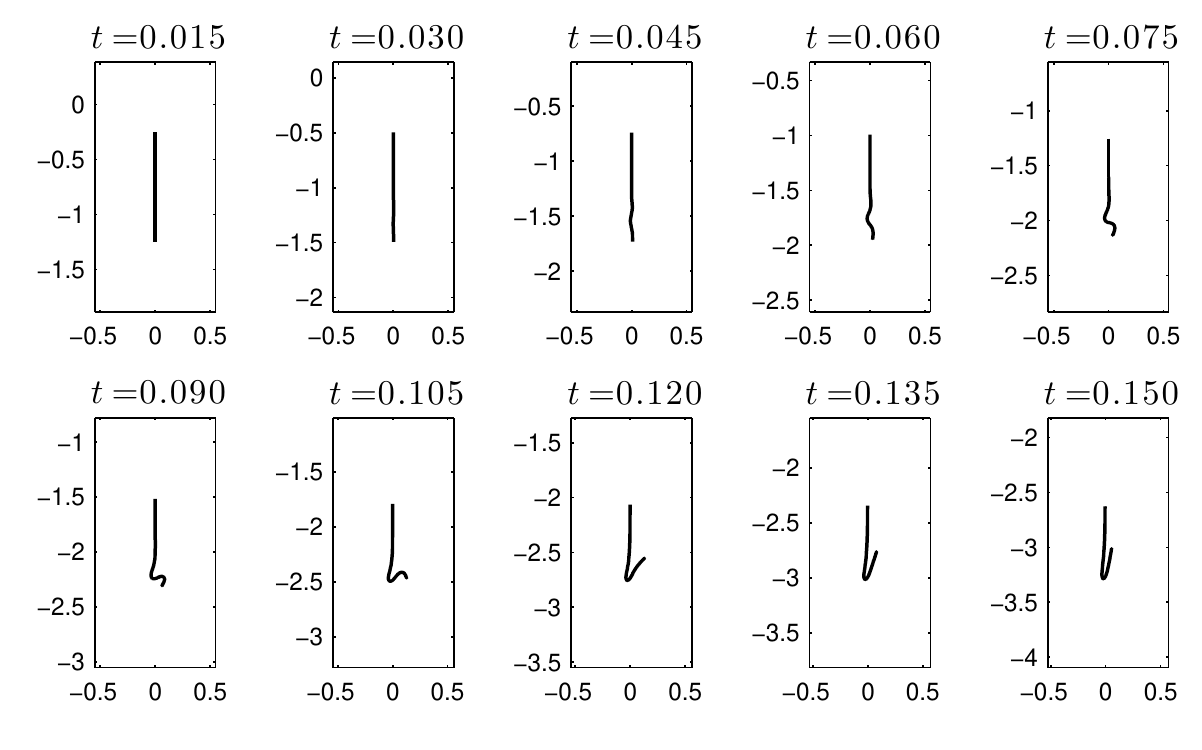}
\caption{Substantial buckling is observed in simulations for $\beta=6.25\times 10^{-5}$ and $B(s)=1$. The filament is initially placed with its trailing end at the origin.}
\label{Figure9}
\end{figure}

Choosing the spheroidal filament profile $r(s)=2\sqrt{s(1-s)}$, so that $c(s)=\log(1/\e^2)$, and setting $B(s)=1$ as before, considerable buckling is observed in the full simulations for sufficiently small values of the elasto-gravitation number. Figures \ref{Figure8} and \ref{Figure9} show time sequences of filaments buckling with $\beta=10^{-4}$ and $\beta=6.25\times 10^{-5}$, respectively (see also supplementary movie 3, which shows similar sequences with three values of $\beta$).  In both cases, an initial transverse perturbation of $10^{-4}\cos(4\upi s)$ is imposed along the entire filament length and is found to amplify and lead to the observed dynamics. Two points are to be noted here, both of which we analyze in further detail in the following sections. First, it can be seen that the buckling instability only occurs in the leading half of the filament, whereas perturbations are observed to decay in the trailing half. This is consistent with the aforementioned argument that the negative (compressive) tension in the leading half drives this instability. Second, perturbations are found to propagate upward in the form of traveling waves in the body frame, eventually dying out once they reach the trailing half. Beyond the times shown in the figures, the filament undergoes substantial bending where the curvature becomes so large that the linearized Euler-Bernoulli formulation used in our model may no longer accurately describe the filament elastodynamics. Additionally, excluded volume effects are expected to come into play as the filament nears itself in the later stages of buckling, though we do not account for direct steric interactions in our simulations.

\subsection{Linear stability analysis and buckling criterion}
For sufficiently small values of the elasto-gravitation number, there is no timescale separation between elastic relaxation and sedimentation. We recall the fluid force per unit length,
\begin{gather}
\v{f}(s)=-(T(s)\v{x}_s)_{s}+\beta(B(s)\v{x}_{ss})_{ss}-\v{F}_g(s).
\end{gather}
As in the simulations just described, we choose the filament profile $r(s)=2\sqrt{s(1-s)}$ and the distributed gravitational forcing $\v{F}_g(s)=F_g(s)\v{\hat{y}}=-6s(1-s)\v{\hat{y}}$. Once again we will consider the corresponding bending stiffness profile, $B(s)=r(s)^4$, but we also study the case $B(s)=1$ for the sake of comparison with the computations.

Consider a straightened filament sedimenting along the $-\v{\hat{y}}$ direction, whose centerline position is expressed as $\v{x}(s,t)=-(s-1/2+Ut)\v{\hat{y}}$, with $U$ the constant sedimentation speed. Then equation~\eqref{x_t}, along with the boundary conditions \eqref{bcs2}, may be written as
\begin{gather}
U=2(c-1)(T_s-F_g)+2S[T_s-F_g],\label{tension_base}\\
T(0)=T(1)=0,\label{tension_basebc}
\end{gather}
where $S[\cdot]$ is the integral operator defined in \eqref{S[]} which is diagonalized under the Legendre polynomial basis as discussed in \S 4.1. Therefore, upon multiplication of \eqref{tension_base} by Legendre polynomials and integrating on $s\in[0,1]$ we see that $T_s(s)-F_g(s)=1$, and that the sedimentation speed is given by $U=2(c-1)$. In addition, noting \eqref{tension_basebc}, we find that the tension along the filament is given by
\begin{gather}
T(s)=T_0(s)=s(1-s)(1-2s).\label{T0(s)}
\end{gather}
We will refer to the straightened filament conformation with sedimentation speed $U=2(c-1)$ and tension $T_0(s)$ as the base state for the analysis to come. Importantly, due to the spatial variation in the gravitational potential, we observe that the tension in the base state is positive for $s\in(0,1/2)$, but negative for $s\in(1/2,1)$. Hence, while the trailing half of the filament experiences a tension, the leading half of the filament (in the direction of sedimentation) experiences a compression. Buckling, therefore, is to be expected in a certain range of $\beta$, but in a non-uniform fashion along the filament backbone (see figures~\ref{Figure8} and \ref{Figure9}).

\subsubsection{Dynamics of filament perturbations\label{sec:fourier}}

We now perform a classical linear stability analysis on the filament by perturbing the filament position in a plane perpendicular to gravity. Assume that the filament position is given by
\begin{gather}
\v{x}(s,t)=-(s-1/2+U t)\v{\hat{y}}+\varepsilon u(s,t)\v{\hat{x}}+O(\varepsilon^2), \label{eq:basestatex}
\end{gather}
with $\varepsilon \ll 1$. By a symmetry argument (taking $\varepsilon\rightarrow -\varepsilon$), it is apparent that there can be no variation in the vertical component of the filament velocity (either in the sedimentation speed or varying spatially along the filament), so the speed $U$ in equation~\eqref{eq:basestatex} is that from the leading-order calculation, $U=2(c-1)$. Performing a regular expansion of the tension about the base state for small $\varepsilon$, we also write
\begin{gather}
T(s)=T_0(s)+\varepsilon T_1(s)+O(\varepsilon^2),
\end{gather}
where $T_1(0)=T_1(1)=0$. Note that $\varepsilon$ at the outset has no relationship to the body aspect ratio $\e$. Inserting these expansions into the $\v{\hat{y}}$ component of the position equation \eqref{x_t}, we have
\begin{gather}
0=2(c-1)T_1'(s)+2S[T_1](s),
\end{gather}
from which we see that the tension does not vary at first order in $\varepsilon$: $T_1(s)=0$. However, upon inspection of the $\v{\hat{x}}$ component of the filament position equation, we find an equation for the dynamics of the perturbation,
\begin{align}
\begin{split}
u_t=&\,(c+1)[(T_0u_s)_s-\beta(B(s)u_{ss})_{ss}]+(c-3)u_s \\
&+S\left[(T_0u_s)_s-\beta(B(s)u_{ss})_{ss}+ u_s\right]+P[u], \label{perturb}
\end{split}
\end{align}
with $P[\cdot]$ defined in \eqref{S[]}.

The analysis of \eqref{perturb} is no longer as simple as an expansion in the Legendre polynomial basis. Instead, we proceed to consider the action of the integral operators on Fourier perturbations of a given high wavenumber $k$. Specifically, for $k \gg 1$, and for points $s$ sufficiently well removed from the filament endpoints, we have
\begin{gather}
S[\mathrm{e}^{\mathrm{i} k s}]\approx -\log\left(\mathrm{e}^{2\gamma} k^2 s(1-s)\right)\mathrm{e}^{\mathrm{i} k s},\\
P[\mathrm{e}^{\mathrm{i} k s}]\approx 2\mathrm{i} k \mathrm{e}^{\mathrm{i} k s},
\end{gather}
as shown in Appendix A, where $\gamma$ is Euler's constant. Hence, for filament perturbations of high wavenumber, the eigenfunctions of $S[\cdot]$ and $P[\cdot]$ are approximately the Fourier basis functions $\mathrm{e}^{\mathrm{i} k s}$. Accordingly, for $k \gg 1$, we may replace $P[u]$ in \eqref{perturb} by $2 u_s$.

As a first approximation justified in Appendix A, we analyze the dynamics of the perturbation in the two halves $s\in(0,1/2)$ (where $T_0(s)>0$) and $s\in(1/2,1)$ (where $T_0(s)<0$) as separate and decoupled. We begin by considering the trailing half of the filament, $s\in(0,1/2)$. While it would be more exhaustive to consider a continuously varying basis for the perturbations, much will be learned by the simpler confinement to a countable
Fourier basis. The Fourier transform and inverse transform pair on this interval are given by
\begin{gather}
u(s,t)=\sum_{k=-\infty}^\infty \hat{u}_k(t)\mathrm{e}^{4\upi \mathrm{i} k s},\,\,\,\,\,\,\,\,\hat{u}_k(t)=2\int_{0}^{1/2} u(s,t) \mathrm{e}^{-4\upi \mathrm{i} k s}\,\mathrm{d}s.
\end{gather}
We also express $(T_0u_s)_s$ and $(Bu_{ss})_{ss}$ in the Fourier basis,
\begin{gather}
 (T_0u_s)_s=\sum_{k=-\infty}^\infty a_k \mathrm{e}^{4\upi \mathrm{i} k s},\ \ \ \ (B(s)u_{ss})_{ss}=\sum_{k=-\infty}^\infty b_k \mathrm{e}^{4\upi \mathrm{i} k s}.
\end{gather}
Using the base-state tension \eqref{T0(s)}, we find
\begin{gather}
a_k=-(\upi k)^2\hat{u}_k+\sum_{m\neq k}\frac{3km(-\mathrm{i}+(m-k)\upi)}{\upi(m-k)^3}\hat{u}_m.
\end{gather}
Also, with $B(s)=1$ we find
\begin{gather}
b_k=(4\upi k)^4\hat{u}_k,
\end{gather}
or with $B(s)=r(s)^4=16s^2(1-s)^2$ we find
\begin{gather}
b_k=8\left(\frac{1}{15}(4\upi k)^4-2\upi \mathrm{i} k\right)\hat{u}_k-\sum_{m\neq k}C(k,m)\hat{u}_m,
\end{gather}
where
\begin{gather}
C(k,m)=128\left[m^4\left(\frac{3+3\mathrm{i}(m-k)\upi+\mathrm{i}(m-k)^3\upi^3}{(m-k)^4}\right)-3m^2(m+k)\frac{1+\mathrm{i}(m-k)\upi}{(m-k)^3}\right].
\end{gather}
Inserting these expressions into \eqref{perturb} returns an equation for the perturbation dynamics in Fourier space,
\begin{gather}
 \hat{u}_k'(t)=(c+1)(a_k-\beta b_k)+(c-1)4\upi \mathrm{i} k\hat{u}_k-2\log(2\upi k \mathrm{e}^{-(1-\gamma)})(a_k-\beta b_k+4\upi \mathrm{i} k\hat{u}_k).\label{ukhatprime}
\end{gather}

The spatially varying tension and bending stiffness lead to the transmission of energy from each wavelength of $u$ to nearby wavelengths through the coefficients $a_k$ and $b_k$. However, consider the case that the filament is seeded with a perturbation with a single wavenumber $k$. For short times, during which the coupling between the Fourier modes can be neglected, we have
\begin{gather}
\hat{u}_k(t)\approx \hat{u}_k(0)\mathrm{e}^{\sigma(k)t}.
\end{gather}
Inserting this ansatz into equation~\eqref{ukhatprime} and neglecting coupling terms, we find the growth rate if $B(s)=1$,
\begin{align}
\begin{split}
\sigma(k)&= \left(c-\log(4(\upi k)^2\mathrm{e}^{2\gamma-3})\right)\left(-\upi^2k^2-\beta(4\upi k)^4\right)
+4\upi \mathrm{i} k(c-\log(4\upi^2 k^2\mathrm{e}^{2\gamma-1}))\\
&\approx -\log\left(\frac{1}{\e^2 k^2}\right)\left(\upi^2k^2+\beta(4\upi k)^4-4\upi \mathrm{i} k\right),
\end{split}
\end{align}
or if $B(s)=16s^2(1-s)^2$,
\begin{align}
\begin{split}
\sigma(k)= &\left(c-\log(4\upi^2 k^2\mathrm{e}^{2\gamma-3})\right)\left(-\upi^2k^2-\frac{8(4\upi k)^4}{15}\beta+16\upi \mathrm{i} k\beta\right)\\
&+4\upi \mathrm{i} k\left(c-\log \left(4\upi^2 k^2\mathrm{e}^{2\gamma-1}\right)\right)\\
\approx& -\log\left(\frac{1}{\e^2 k^2}\right)\left(\upi^2k^2+\frac{8}{15}\beta(4\upi k)^4-4\upi \mathrm{i} k(1+4\beta)\right),
\end{split}
\end{align}
where we have inserted $c=\log(1/\e^2)$ with $\e$ the filament aspect ratio. The growth rate $\sigma(k)$ exhibits rapid damping due to bending rigidity ($\propto -k^4$) as well as damping due to filament tension ($\propto -k^2$). The perturbation is thus expected to return rapidly to its straightened state. The dispersion relation also shows that the perturbation travels as a wave along the filament in the direction opposite gravity with approximate speed $\log(1/\e^2 k^2)$. The approximation clearly breaks down if the filament aspect ratio is on the order of the perturbation wavelength, $\e k=1$, so we assume $\e k \ll 1$.

Meanwhile, in the leading half of the filament, $s\in(1/2,1)$, there is a slight but critical adjustment to the approximations above, as a consequence of the negative tension there. By a similar calculation, we find the short-time growth rates for $B(s)=1$,
\begin{gather}
\sigma(k) \approx \log\left(\frac{1}{\e^2 k^2}\right)\left(\upi^2k^2-\beta(4\upi k)^4+4\upi \mathrm{i} k\right),\label{localdispersion}
\end{gather}
and separately for $B(s)=16s^2(1-s)^2$,
\begin{gather}
\sigma(k)\approx\log\left(\frac{1}{\e^2 k^2}\right)\left(\upi^2k^2-\frac{8}{15}\beta(4\upi k)^4+4\upi \mathrm{i} k(1-4\beta)\right).
\end{gather}
In the leading half, we observe a competition between the effect of tension, which acts to amplify the perturbation exponentially fast, and the effect of bending rigidity, which acts to dampen the system. In the case $B(s)=16s^2(1-s)^2$, the filament is predicted to buckle for wavenumbers smaller than a critical value, $k^*=\sqrt{15}/(16 \upi \sqrt{8\beta})$, and the most unstable wavenumber (corresponding to the largest positive growth rate) is given by $k_m=\sqrt{15/\beta}/(64\upi)$. While arbitrarily small wavenumbers can be supported by a free filament, the critical value of $\beta$ for which at least one wavelength of buckling can be observed $(k^*=1/2)$ is $\beta^*=15/(8(8\upi)^2)\approx 0.0030$ in this case. Once again, the growing perturbation travels as a wave in the body frame in the direction opposite gravity, as observed in the numerical simulations of figures \ref{Figure8} and~\ref{Figure9}.

\begin{figure}
\centering
\includegraphics{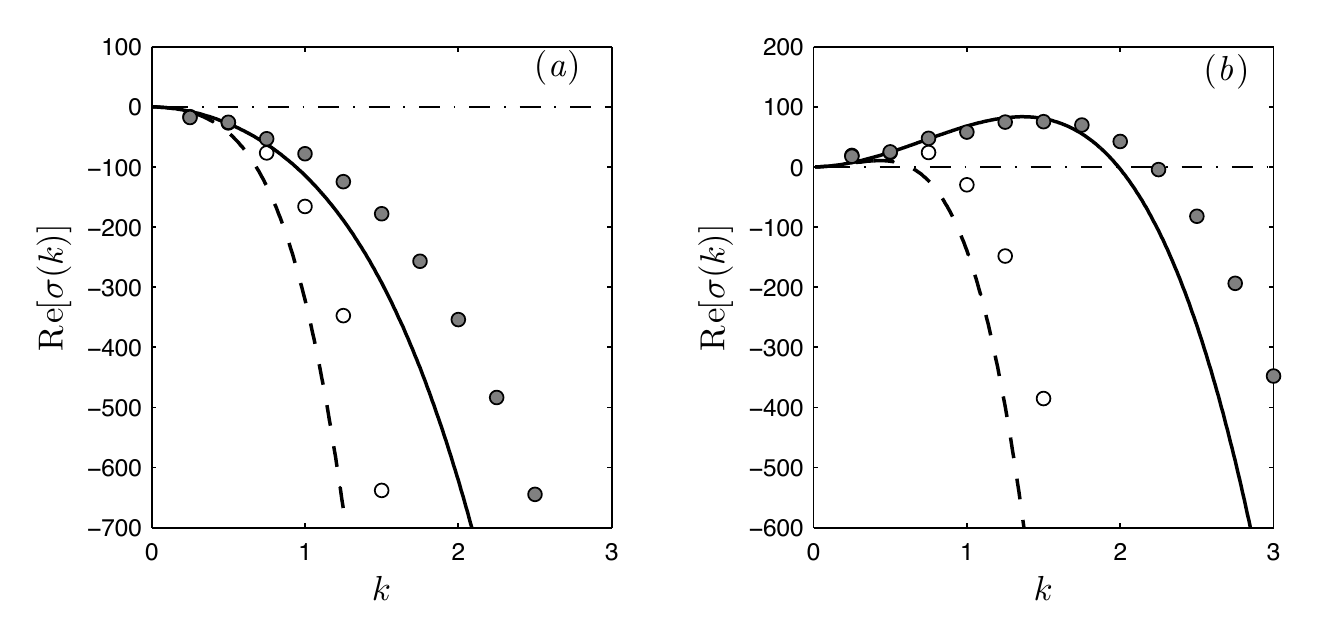}
\caption{Real part of the growth rate $\sigma(k)$, for two different values of $\beta$ in the (\textit{a}) trailing and (\textit{b}) leading halves of the filament. The lines are theoretical predictions and symbols follow from simulations. The solid lines and filled circles correspond to $\beta=10^{-4}$ while the dashed line and open symbols correspond to $\beta=10^{-3}$.}
\label{Figure10}
\end{figure}

To quantitatively compare the analytical predictions with the full numerical results, we perform simulations in which an initially straight and vertically aligned filament is weakly perturbed at a given wavenumber $k$ across its entire length at $t=0$: $u(s,0)=10^{-4}\cos(4 \upi k s)$, and we set $B(s)=1$. The effective growth rates of such perturbations in the linear regime are extracted numerically and are compared to the predictions of the linear analysis in figure \ref{Figure10} for two different values of $\beta$. In agreement with the theoretical predictions, the trailing half of the filament is always found to be stable to single wavenumber perturbations, while the leading half is unstable over a finite range of wavenumbers where the competition between compressive tension and elasticity is favorable for buckling to occur. The numerical results and theoretical growth rates follow similar trends, though damping is always found to be smaller in the simulations. This systematic shift, which becomes more apparent for larger values of $k$, may be due to the coupling between the two halves, the coupling between modes, and the filament boundary conditions, which have all been neglected for this first approximation. In particular, to achieve the above estimation the assumption was made that the region of interest is well-separated from the filament endpoints, whereas the instability observed in the simulations is dominant near the leading tip of the filament. Perturbations in the trailing half do indeed decay as predicted, in the form of upward traveling waves.

As discussed earlier and illustrated in figure~\ref{Figure7}(\textit{b}), the tension in the leading half of a filament with uniform thickness is also negative in the straightened (base) state, but due instead to nonlocal hydrodynamic interactions. An approximation of the tension accurate to $O((\log 1/\epsilon)^{-2})$ for this case is derived in Appendix B.

\subsection{Linear eigenmodes of the local theory}

The growth rate derived above was based on the short-time behavior of a Fourier perturbation of wavenumber $k$, where we neglected the couplings between wavenumbers and assumed that the stability of the leading and trailing halves of the filament could be analyzed independently. Fourier modes, however, are not exact eigenfunctions of the linearized equation \eqref{perturb}, in particular near the filament endpoints, which may explain the quantitative discrepancies we observed between the theoretical and numerical growth rates in figure~\ref{Figure10}. A different approach which is semi-analytical consists in solving for the exact eigenfunctions of the problem that are valid along the entire length of the filament, as was previously done by \cite{Young07} and \cite{gkss12} for the buckling of elastic fibers in extensional flows near hyperbolic stagnation points.

\begin{figure}
\centering
\includegraphics{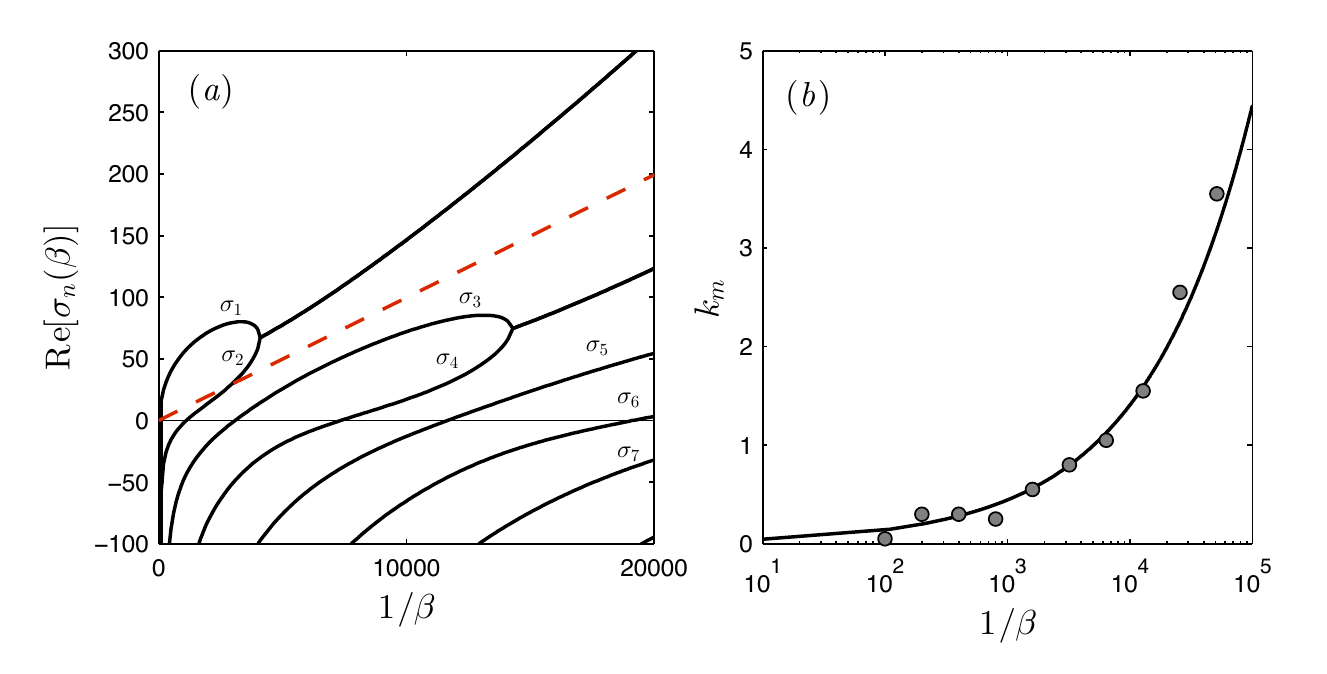}
\caption{(Colour online) (\textit{a}) Real parts of the largest few eigenvalues $\sigma_n$ as a function of $1/\beta$. Also shown with a dashed line is the maximum growth rate as predicted by the local dispersion relation of \eqref{localdispersion}. (\textit{b}) The wavenumber corresponding to the largest discrete Fourier component of the most unstable eigenfunction (circles) compared against the wavenumber with the largest growth rate in the local dispersion relation (solid line).}
\label{Figure11}
\end{figure}

We determine numerically the linear eigenmodes of the problem in the case where the nonlocal contribution is neglected. Keeping only the local contribution in equation~\eqref{perturb} and setting $B(s)=1$ for simplicity, the linearized equation for the amplitude of the shape fluctuations becomes
\begin{equation}\label{eq:linearized}
u_t(s,t)=(c+1)[(T_0 u_s)_s-\beta u_{ssss}]+(c-3)u_s,
\end{equation}
with $T_0(s)$ the base-state tension from equation \eqref{T0(s)}. In the linear regime, we seek exponentially growing solutions of the form $u(s,t)=\varphi_n(s)\mathrm{e}^{\sigma_n t}$, where the eigenfunctions $\varphi_n(s)$ satisfy
\begin{equation}\label{eq:evalproblem}
\sigma_n \varphi_n=2[c-1-3(c+1)s(1-s)](\varphi_n)_s+(c+1)[s(1-s)(1-2s)(\varphi_n)_{ss}-\beta(\varphi_n)_{ssss}].
\end{equation}
Given that $B(s)=1$ and that the tension profile vanishes at the ends in the linear regime, the boundary conditions \eqref{bcs1}--\eqref{bcs2}  simply become
\begin{equation}\label{eq:boundaries}
\varphi_n''(0)=\varphi_n''(1)=\varphi_n'''(0)=\varphi_n'''(1)=0.
\end{equation}
Equation (\ref{eq:evalproblem}) is an eigenvalue problem for the mode shapes $\varphi_{n}(s)$, with corresponding eigenvalues $\sigma_{n}$, whose real parts define the growth rates. We solve the equation numerically using a second-order accurate finite-difference discretization, which yields a countable set of eigenfunctions and eigenvalues. The eigenvalues are ordered by decreasing values of the growth rate, $\mathrm{Re}(\sigma_{1})\ge\mathrm{Re}(\sigma_{2})\ge ...$ The largest growth rates, which correspond to the most unstable modes, are plotted as functions of $1/\beta$ in figure~\ref{Figure11}(\textit{a}), where we observe that an increasing number of modes become unstable with increasing filament flexibility (decreasing $\beta$). Nevertheless, we find that the first mode with eigenvalue $\sigma_{1}$ always remains the most unstable (though it merges with the second mode when $\beta\lesssim 2.5\times 10^{-4} $ as we discuss below), and the maximum growth rate is found to compare favorably with the results of the Fourier analysis of \S \ref{sec:fourier}.

\begin{figure}
\centering
\includegraphics{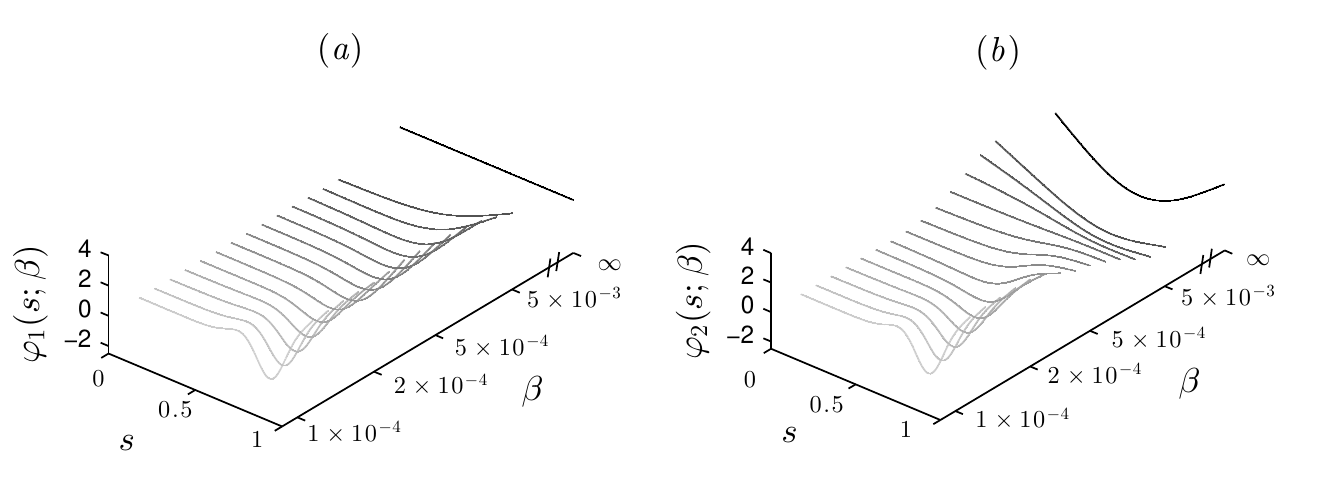}
\caption{Eigenfunctions $\varphi_1$ (\textit{a}) and $\varphi_2$ (\textit{b}) for $\beta$ in the range $1\times10^{-4}-5\times10^{-3}$. In the limit of $\beta\rightarrow \infty$, the eigenfunctions are biharmonic functions, and the shapes progressively become less symmetric for more flexible filaments. Note that $\varphi_1$ and $\varphi_2$ become identical below $\beta\approx 2.5\times10^{-4}$.}
\label{Figure12}
\end{figure}

The shapes of the eigenmodes are illustrated in figure~\ref{Figure12}, which shows the two most unstable eigenfunctions $\varphi_{1}(s)$ and $\varphi_{2}(s)$ for values of the elasto-gravitation number $\beta$ in the range $1\times10^{-4} - 5\times10^{-3}$. Note that in the limit of $\beta\rightarrow \infty$ (stiff filaments), the eigenfunctions are simply eigenfunctions of the biharmonic operator, but these lose symmetry with decreasing $\beta$ as the filament becomes more flexible and hence susceptible to buckling in a nonuniform fashion as we have described. The modal stability in figure~\ref{Figure11}(\textit{a}) shows real parts of eigenvalues merging as $\beta$ decreases: this is seen here as the shapes of eigenfunctions $\varphi_{1,2}$  become identical below $\beta\approx2.5\times10^{-4}$ when the two eigenvalues $\sigma_{1,2}$ become complex conjugates. Additionally, $\varphi_{1,2}$ remains the most unstable buckling mode as $\beta$ decreases further and more complicated shapes involving higher wavenumbers arise. We see that the eigenvalues for the problem on the whole interval roughly agree with the predicted growth rates from the previous section as a consequence of the most unstable eigenfunction only taking significant values on the leading half of the filament, $s\in(1/2,1)$. We note a striking similarity between the linearly unstable eigenmodes calculated here for $n=1,2$ and the finite-amplitude buckled shapes observed in the nonlinear numerical simulations of figures~\ref{Figure8} and \ref{Figure9}.

The increasing wavenumber content of the unstable eigenfunctions with increasing flexibility is consistent with the widening range of unstable wavenumbers predicted by the Fourier analysis of \S \ref{sec:fourier}. To compare both results more quantitatively, we project the leading half $s\in(1/2,1)$ of $\varphi_{1}(s)$ onto a discrete Fourier cosine basis $\cos(4\upi k s)$ and compare the wavenumber $k_{m}$ with the dominant projection to the most unstable wavenumber predicted by the dispersion relation of \S \ref{sec:fourier}. As shown in figure~\ref{Figure11}(\textit{b}), both values match closely, which further corroborates the use of a countable Fourier basis for the stability analysis.

\section{Conclusion}

In this paper we investigated some of the fundamental dynamics of a single flexible filament as it sediments in a viscous fluid. The competition between elastic forces and viscous forces induced by gravity was characterized by a dimensionless quantity that we termed the elasto-gravitation number, $\beta$. We first considered the weakly flexible regime, where the filament is nearly rigid, and using a multiple-scale analysis found a self-similar scaling of the filament shape with an amplitude dependent upon the body orientation. Equilibrium shapes and trajectories were then analyzed in this regime, and we gave predictions for the dynamics of clouds of multiple (non-interacting) filaments. By comparing against full numerical simulations, the analytical predictions were found to be accurate for elasto-gravitation numbers down to $\beta\approx 0.01$ in the case of spheroidal filaments with thickness profile $r(s)=2\sqrt{s(1-s)}$. A similar analysis was provided in Appendix B for the shapes, velocities, and rotation rates of filaments with uniform thickness.

We then turned our attention to the buckling of a very flexible filament sedimenting along its long axis, which can occur for sufficiently small elasto-gravitation numbers. While arbitrarily small wavenumbers can be supported by a free-filament, the critical value of $\beta$ for which one wavelength of buckling can be observed was found to be $\beta^*=15/(8(8\upi)^2)\approx 0.0030$ in the case of the spheroidal filament with $B(s)=16s^2(1-s^2)$. Two approaches were taken to study the most unstable wavelength perturbations and their growth rates. In the first approach, we assumed highly oscillatory perturbations so that the nonlocal integral operator could be handled analytically and the filament endpoints did not play a role. In the second, we solved numerically an eigenvalue problem for the most unstable eigenmodes on the full interval. Both approaches yielded predictions that lay in agreement, which in turn matched very closely with the results of full numerical simulations.

Future work might consider the sedimentation of many flexible bodies from experimental, numerical, and analytical perspectives. Issues such as filament entanglement may dominate the dynamics in systems with sufficiently close-packed flexible filaments. As a final observation, we note that the filament shapes, trajectories, and instabilities studied in this paper can be interpreted equivalently as those of a positively buoyant flexible filament rising against gravity.

\begin{acknowledgments}
We thank Arthur Evans and Randy Ewoldt for helpful conversations, and gratefully acknowledge funding from an FMC Educational Fund fellowship (H.M.) and from NSF CAREER Grant CBET-1150590 (D.S.).
\end{acknowledgments}
\vspace{.1in}
Supplementary movies of weakly flexible sedimenting filaments and of the buckling of highly flexible filaments are available at journals.cambridge.org/ßm.

\appendix

\section{Reduction of integral operators for high wavenumber perturbations}
When analyzing the instability of the two halves of the filament in \S 5, we use that the Fourier basis functions approximately diagonalize the integral operators in \eqref{S[]} for large wavenumbers, $k \gg1$, and we also decouple the integral operators into operations on the two halves of the filament separately. We now justify both approximations.

Consider a point $s\in [\Delta, 1-\Delta]$, with $\Delta>0$. Then for $\lambda\geq 1$, with $k\gg 1$ such that $k\Delta$ is sufficiently large, the action of the integral operators in \eqref{K[f]} on the Fourier basis functions yield
\begin{gather}
S[\mathrm{e}^{\mathrm{i}\lambda ks}]\approx (-2\gamma-2\log(\lambda k)-\log(s(1-s)))\mathrm{e}^{\mathrm{i}\lambda ks},\label{approxs}\\
P[\mathrm{e}^{\mathrm{i}\lambda ks}]\approx 2\mathrm{i}(\lambda k)\mathrm{e}^{\mathrm{i}\lambda ks},
\end{gather}
where $\gamma\approx 0.577$ is Euler's constant. To show this, we simply consider a change of variables, $\xi=\lambda ks'$ so that
\begin{gather}
S[\mathrm{e}^{\mathrm{i} \lambda ks}](s)=\Lambda(k,s)\mathrm{e}^{\mathrm{i} \lambda ks},
\end{gather}
with
\begin{align}
\begin{split}
\Lambda(k, s)&=\int_{-\lambda k s}^{\lambda k(1-s)}\frac{\mathrm{e}^{\mathrm{i}\xi}-1}{|\xi|}\,\mathrm{d}\xi\\
&\approx 2\int_0^1\frac{\cos\xi-1}{\xi}\,\mathrm{d}\xi+2\int_1^{+\infty}
\frac{\cos(\xi)}{\xi}\,\mathrm{d}\xi-2\log(\lambda k)-\log(s(1-s)),
\end{split}
\end{align}
which gives the desired result. The same change of variables gives
\begin{gather}
P[\mathrm{e}^{\mathrm{i}\lambda k s}]
=\lambda k \mathrm{e}^{\mathrm{i}\lambda k s}\int_{-\lambda k s}^{\lambda k(1-s)}\frac{\mathrm{e}^{\mathrm{i}\xi}-1-\mathrm{i}\xi\mathrm{e}^{\mathrm{i}\xi}}{\xi|\xi|}\,\mathrm{d}\xi
\approx 2\mathrm{i}\lambda k \mathrm{e}^{\mathrm{i}\lambda k s}\int_0^{+\infty}\frac{\sin\xi-\xi\cos\xi}{\xi^2}\,\mathrm{d}\xi=2\mathrm{i}\lambda k \mathrm{e}^{\mathrm{i}\lambda k s}.
\end{gather}

Moreover, the main contribution of $S[\mathrm{e}^{\mathrm{i}\lambda ks}](s)$ and $P[\mathrm{e}^{\mathrm{i}\lambda ks}](s)$ comes from the neighborhood of $s$. Specifically, for an interval $I\subset [0,1]$ we define
\begin{gather}
S_{I}[\mathrm{e}^{\mathrm{i}\lambda ks}]=\int_{I}\frac{\mathrm{e}^{\mathrm{i}\lambda ks'}-\mathrm{e}^{\mathrm{i}\lambda ks}}{|s'-s|}\,\mathrm{d}s',\\
P_{I}[\mathrm{e}^{\mathrm{i}\lambda ks}]=\int_{I}\frac{(\mathrm{e}^{\mathrm{i}\lambda ks'}-\mathrm{e}^{\mathrm{i}\lambda ks})/(s'-s)-\mathrm{i}\lambda k \mathrm{e}^{\mathrm{i}\lambda ks'}}{|s'-s|}\,\mathrm{d}s',
\end{gather}
and we will show that
\begin{gather}
 S_{[0, 1/2]}[\mathrm{e}^{\mathrm{i}\lambda ks}]\approx
(-2\gamma-2\log(\lambda k)-\log(s(1/2-s)))\mathrm{e}^{\mathrm{i}\lambda ks}\label{S1a},\\
P_{[0, 1/2]}[\mathrm{e}^{\mathrm{i}\lambda ks}]\approx 2\mathrm{i}\lambda k\mathrm{e}^{\mathrm{i}\lambda ks},
\end{gather}
when $s\in (\Delta, 1/2-\Delta)$, and
\begin{gather}
S_{[1/2, 1]}[\mathrm{e}^{\mathrm{i}\lambda ks}]\approx
(-2\gamma-2\log(\lambda k)-\log((s-1/2)(1-s)))\mathrm{e}^{\mathrm{i}\lambda ks},\label{S1b}\\
P_{[1/2, 1]}[\mathrm{e}^{\mathrm{i}\lambda ks}]\approx 2\mathrm{i}\lambda k\mathrm{e}^{\mathrm{i}\lambda ks},
\end{gather}
when $s\in (1/2+\Delta, 1-\Delta)$. To see this, let
\begin{gather}
S_{[0,1/2]}[\mathrm{e}^{\mathrm{i}\lambda k s}]=I_1\mathrm{e}^{\mathrm{i}ks},\\
S_{[1/2,1]}[\mathrm{e}^{\mathrm{i}\lambda k s}]=I_2\mathrm{e}^{\mathrm{i}ks}.
\end{gather}
and
\begin{gather}
P[\mathrm{e}^{i\lambda k s}]=P_{[0, 1/2]}[\mathrm{e}^{\mathrm{i}\lambda k s}]+P_{[1/2, 1]}[\mathrm{e}^{\mathrm{i}\lambda k s}]
\end{gather}

We have when $s\in[\Delta, 1/2-\Delta]$, with $k \gg 1$,
\begin{gather}
I_1\approx-2\gamma-2\log(\lambda k)-\log(s(1/2-s)),\\
I_2\approx\log((1-s)/(1/2-s)).
\end{gather}
and
\begin{gather}
P_{[0,1/2]}/(\lambda k)\approx 2\mathrm{i}\mathrm{e}^{\mathrm{i}\lambda k s},\\
P_{[1/2, 1]}/(\lambda k) \approx 0.
\end{gather}
We observe on this interval that $I_1$ dominates $I_2$ and the first integral for $P$ dominates the second.
Similar computation yields for $s\in [\Delta, 1/2-\Delta]$. A further approximation, leading to \eqref{ukhatprime}, is obtained using
\begin{gather}
2\int_{0}^{1/2}\log(s(1/2-s))\,\mathrm{d}s=2\int_{1/2}^{1}\log((s-1/2)(1-s))\,\mathrm{d}s
=-2-2\log 2.
\end{gather}
We now simply replace the term $\log(s(1/2-s))$ in \eqref{S1a} and the term $\log((s-1/2)(1-s))$ in \eqref{S1b} by $-2-2\log 2$. In addition, we see that $P[u]\approx 2u_s \approx P_I[u]$ for high wavenumber perturbations.

\section{A filament of uniform thickness: bending and buckling by nonlocal hydrodynamic interactions}

In both \S 4 and \S5 we considered the filament profile $r(s)=2\sqrt{s(1-s)}$ for mathematical convenience. The equilibrium shapes found in figure~\ref{Figure3}, for instance, were a consequence of variations in gravitational potential and viscous drag along the filament length. However, filaments of uniform thickness, $r(s)=1$, are expected to result in qualitatively similar shapes but instead as a consequence of a secondary effect, namely by nonlocal hydrodynamic interactions. As illustrated in figure~\ref{Figure1}, the central segments of the filament experience a stronger disturbance flow and will sediment faster than segments nearer to the filament ends. An accompanying reorientation is also to be expected. This case was considered by \cite{xn94}. Similarly, as illustrated in figure~\ref{Figure7}, a sufficiently flexible filament of uniform thickness is also expected to buckle when sedimenting along its long axis. While the effects due to variations in the filament thickness are $O(1)$, the effects due to nonlocal hydrodynamic interactions will be shown to be considerably smaller, $O(\ln(1/\epsilon)^{-1})$.

In both regimes (weakly and highly flexible filaments), the leading-order hydrodynamic interaction appears in the equations of motion through the spatially varying function $c(s)$ in \eqref{lambda[f]}, while a higher-order correction is given by the nonlocal integration of \eqref{K[f]}, which itself is now made considerably more challenging analytically. We will now proceed to derive the shapes and velocities of filaments with uniform thickness in the weakly flexible regime, as well as the base-state tension for sedimentation along the filament's long axis.

\subsection{Reorientation of a weakly flexible filament}
Choosing $r(s)=1$, we have a uniform distribution of gravitational potential and bending stiffness, $F_g(s)=-1$ and $B(s)=1$. Equations \eqref{eq:leadingorder1} and \eqref{eq:leadingorder2} then become:
\begin{gather}
U^{(0)}=2(c(s)-1)\left[T^{(0)}_s+\cos\theta^{(0)}\right]+2S\left[T^{(0)}_s\right],\label{eq:leaduni1}\\
V^{(0)}+(s-1/2)\theta^{(0)}_t=-(c(s)+1)\left[u_{ssss}+\sin\theta^{(0)}\right]-S\left[u_{ssss}\right],\label{eq:leaduni2}
\end{gather}
where now $c(s)=c_0+\log(4s(1-s))$, with $c_0=\log(1/\e^2) \gg 1$. We pursue approximate expressions at leading order in the small number $1/c_0$. It is straightforward to show that $U^{(0)}=O(c_0)$, $V^{(0)}=O(c_0)$, $T^{(0)}=O(c_0^{-1})$, and $u=O(c_0^{-1})$. Therefore we assume the following series expansions, 
\begin{gather}
V^{(0)}=\sum_{n=0}^{+\infty}V_nc_0^{1-n},\ \ 
\theta^{(0)}=\sum_{n=0}^{+\infty}\theta_nc_0^{-n}, \  \
\sin\theta^{(0)}=\sum_{n=0}^{+\infty}a_nc_0^{-n}, \ \ 
u(s)=\sum_{n=0}^{+\infty}u_n(s)c_0^{-n-1}.
\end{gather}

Upon insertion into \eqref{eq:leaduni2}, we obtain:
\begin{gather}
V_1+(s-1/2)\partial_t\theta_{0}=-[1+\ln(4s(1-s))]a_0-u''''_{0},\\
V_{n+1}+(s-1/2)\partial_t\theta_{n}=-[1+\ln(4s(1-s))](a_n+u''''_{n-1})-u''''_{n}-S[u''''_{n-1}].
\end{gather}
Multiplying by $(s-1/2)$ and integrating, and using the boundary conditions $u'''_{n}(0)=u'''_{n}(1)=0$, $u''_{n}(0)=u''_{n}(1)=0$, we find that $u_{n}(s)$ is symmetric about $1/2$ for all $n$ and $\partial_t \theta_{n}=0$ by induction. Therefore, there is no rotation at leading order in $1/\beta$, $\theta_t^{(0)}=0$.

\begin{figure}
\begin{center}
\includegraphics[width=2.8in]{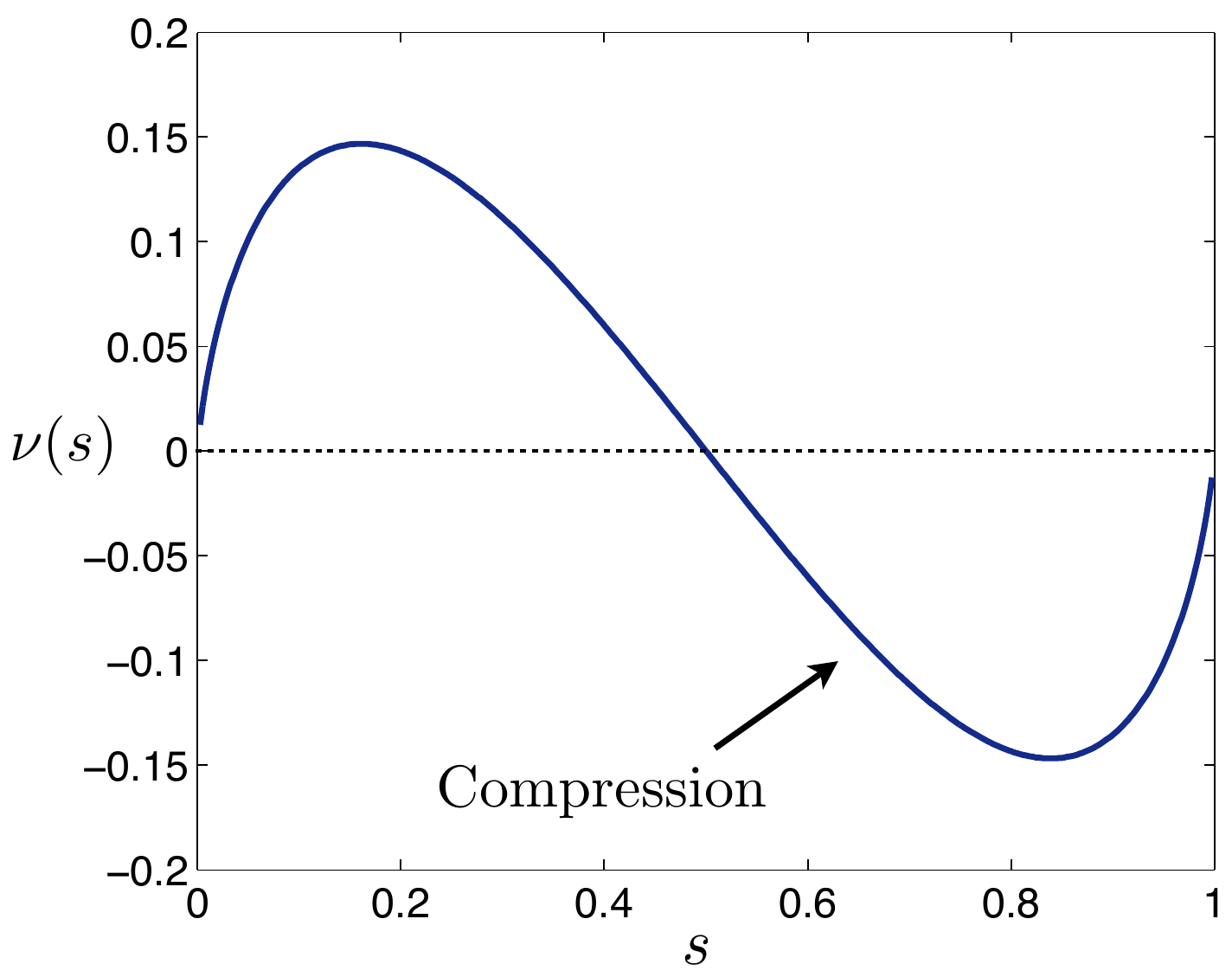}
\caption{The tension profile along a straight filament of uniform thickness due to nonlocal hydrodynamic interactions, from equation~\eqref{T0B}, whose amplitude is proportional to $\cos \theta^{(0)}$. Once again, buckling is possible in the leading half of the filament, where $s\in(1/2,1)$ (see figure~\ref{Figure7}).}
\label{Figure13}
\end{center}
\end{figure}

For the sake of convenience we again let $\zeta=s-1/2$. Then, defining the series expansions
\begin{gather*}
U^{(0)}=c_0U_0+U_1+O(c_0^{-1}),\ \ \ \ V^{(0)}=c_0V_0+V_1+O(c_0^{-1}),\\
T^{(0)}=c_0^{-1}T_0+O(c_0^{-2}),\ \ \ u=c_0^{-1}u_0+O(c_0^{-2}),\\
\theta^{(0)}=\theta_0+c_0^{-1}\theta_1+O(c_0^{-2}),
\end{gather*}
we find
\begin{gather}
U_0=2\cos\theta_0,\ \ \ V_0=-\sin\theta_0,\\
U_1+2\theta_1\sin\theta_0=2[\ln(1-4\zeta^2)-1]\cos\theta_0+2T_0',\\
V_1+\theta_1\cos\theta_0=-u''''_{0}-[\ln(1-4\zeta^2)+1]\sin\theta_0.
\end{gather}
Using the boundary conditions on $T_0(s)$ and $u_0(s)$, we find:
\begin{gather}
T_0'+[\ln(1-4\zeta^2)+2(1-\log 2)]\cos\theta_0=0,\\
u''''_{0}+[\ln(1-4\zeta^2)+2(1-\log 2)]\sin\theta_0=0,
\end{gather}
leading to the leading order tension profile,
\begin{gather}
T_0(\zeta)=\left[2\log(2)\zeta-\Big(\zeta-\frac{1}{2}\Big)\log(1-2\zeta)
-\Big(\zeta+\frac{1}{2}\Big)\log(1+2\zeta)\right]\cos\theta_0,
\end{gather}
and the leading-order filament deflection profile,
\begin{multline}
u_0(\zeta)=\frac{1}{24}\Big[-(\zeta-1/2)^4\log(1-2\zeta)-(\zeta+1/2)^4
\log(1+2\zeta)\\
+\left(\frac{13}{6}+2\log 2\right)\zeta^4
+\frac{1}{4}\left(12\log 2+1\right)\zeta^2\Big]\sin\theta_0.
\end{multline}
This shape is the same as that derived by \cite{xn94} when $\theta_0=\pi/2$, though with the correction of a small typo. Reinserting $\zeta=s-1/2$, we have that 
\begin{gather}
T^{(0)}(s)=c_0^{-1}\nu(s)\cos\theta^{(0)}+O(c_0^{-2}),\label{T0appendix}\\
u(s)=c_0^{-1}h(s)\sin\theta^{(0)}+O(c_0^{-2}),
\end{gather}
with
\begin{gather}
\nu(s)=(1-s)\log(1-s)-s\log s,\label{nu}
\end{gather}
and
\begin{multline}
h(s)=-\frac{1}{24}\Big[(s-1)^4\log(2-2s)-s^4\log(2s)\\
+\left(\frac{13}{6}+2\log 2\right)(s-1/2)^4
+\left(3\log 2+\frac{1}{4}\right)(s-1/2)^2\Big].
\end{multline}
The tension is therefore zero at leading order when the filament is sedimenting perpendicular to its long axis, but otherwise varies along the filament. A plot of $\nu(s)$ is shown in figure~\ref{Figure13}. Consequently, \eqref{V1} may now be written as:
\begin{align}
\begin{split}
V^{(1)}+&(s-1/2)\left(\theta^{(0)}_\tau+\theta^{(1)}_t\right)=(c(s)+1)\left[(T^{(0)}u_s)_s-u''''_{1}-\theta^{(1)}\cos\theta^{(0)}\right]\\
&+(c(s)-3)(T_s^{(0)}+\cos\theta^{(0)})u_s+S\left[(T^{(0)}u_s)_s-u''''_{1}-\theta^{(1)}\cos\theta^{(0)}\right]\\
&+\int_0^1\frac{\Delta u(T_s^{(0)}(s')+\cos\theta^{(0)})-u_s(s)(T_s^{(0)}(s)+\cos\theta^{(0)})}{|s-s'|}\,\mathrm{d}s'.
\end{split}
\end{align}

Multiplying by $(s-1/2)$ and integrating, and imposing $\theta_t^{(1)}=0$ to remove the secular term, we have:
\begin{gather}
\theta_{\tau}^{(0)}= 6[J_0+c_0^{-1}(J_1+J_2+J_3)]\sin(2\theta^{(0)})+O(c_0^{-2}),
\end{gather}
where
\begin{gather}
J_0=\int_0^1(s-1/2)h_s\,\mathrm{d}s,\\
J_1=\int_0^1(s-1/2)(\nu h_s)_s\,\mathrm{d}s,\\
J_2=\int_0^1(s-1/2)\nu_sh_s\,\mathrm{d}s,\\
J_3=\int_0^1(s-1/2)[\log(4s(1-s))-3]h_s\,\mathrm{d}s.
\end{gather}
Inserting $\nu(s)$ and $h(s)$ from above, the resulting rotation rate in terms of the single time $t$ is given by
\begin{gather}
\theta_t=\frac{1}{\beta}\left(\frac{7}{400}+\log(1/\e^2)^{-1}\frac{749-150 \pi ^2+315 \log(2)}{9000}\right)\sin(2\theta)+O(\log(1/\e^2)^{-2},\beta^{-2})\nonumber \\
=\frac{1}{\beta}\left(0.003-0.057\log(1/\e^2)^{-1}\right)\sin(2\theta)+O(\log(1/\e^2)^{-2},\beta^{-2}).
\end{gather}
The result is physical for $\e < 2 \exp(107/45-10 \pi^2/21)\approx 0.196$. This expression may be compared to that for the spheroidal filament shown in \eqref{theta_0eqn2}. The timescale for reorientation is now significantly longer than that found for the spheroidal filament. 

\subsection{Compression of a uniform flexible filament}
As illustrated in figure~\ref{Figure7}(\textit{b}), the tension in the leading half of a filament with uniform thickness, sedimenting along its long axis, is still expected to be negative in the straightened state due to nonlocal hydrodynamic interactions. The base-state tension and sedimentation speed was already derived for this case in the previous section as the special case $\theta^{(0)}=0$, from which we find
\begin{gather}
U=2c_0+2(2\log(2)-3)+O(c_0^{-1}),\\
T(s)=c_0^{-1}\nu(s)+O(c_0^{-2}),\label{T0B}
\end{gather}
with $c_0=\log(1/\e^2)$, and $\nu(s)$ defined in \eqref{nu} and plotted in figure~\ref{Figure13}. We observe that $T(s)>0$ in the trailing half of the filament, $s<1/2$, and that $T(s)<0$ in the leading half of the filament, $s>1/2$. Buckling is therefore still possible in the leading half of the filament as a consequence of nonlocal hydrodynamic interactions even if the filament has uniform thickness.

\bibliographystyle{jfm}
\bibliography{Bigbib}

\end{document}